\documentclass[%
 reprint,
superscriptaddress,
 amsmath,amssymb,
 aps]{revtex4-2}

\usepackage{lmodern}

\usepackage[colorlinks=true,linkcolor=red,citecolor=blue,urlcolor=blue]{hyperref}

\usepackage{graphicx} 
\usepackage{amsmath,amssymb,physics}
\usepackage{cleveref}
\usepackage{orcidlink}

\usepackage{xcolor}
\usepackage{times}

\newcommand{\eg}{e.g.,~}
\newcommand{\ie}{i.e.,~}

\begin{document}

\title{Dynamical response of twin stars to perturbations}

\author{Shamim Haque \orcidlink{0000-0001-9335-5713}}
\email{shamims@iiserb.ac.in}
\affiliation{Indian Institute of Science Education and Research Bhopal, Bhopal 462066, India}

\author{Luciano Rezzolla \orcidlink{0000-0002-1330-7103}}
\email{rezzolla@itp.uni-frankfurt.de}  
\affiliation{Institut f\"ur Theoretische Physik, Goethe Universit\"at, 
  Max-von-Laue-Str. 1, 60438 Frankfurt am Main, Germany}
\affiliation{School of Mathematics, Trinity College, Dublin 2, Ireland}
\affiliation{Frankfurt Institute for Advanced Studies, 
  Ruth-Moufang-Str. 1, 60438 Frankfurt am Main, Germany}

\author{Ritam Mallick \orcidlink{0000-0003-2943-6388}}
\email{mallick@iiserb.ac.in}
\affiliation{Indian Institute of Science Education and Research Bhopal, Bhopal 462066, India}

\date{\today}

\begin{abstract}
If a strong first-order phase transition takes place at sufficiently high
rest-mass densities in the equation of state (EOS) modelling compact
stars, a new branch will appear in the mass-radius sequence of stable
equilibria. This branch will be populated by stars comprising a
quark-matter core and a hadronic-matter envelope, \ie hybrid stars, which
represent ``twin-star'' solutions to equilibria having the same mass but
a fully hadronic EOS. While both branches are stable to linear
perturbations, it is unclear which of the twin solutions is the
``favoured'' one, that is, which of the two configurations is expected to
be found in nature. We assess this point by performing a large campaign
of general-relativistic simulations aimed at assessing the response of
compact stars on the two branches to perturbations of various
strength. In this way, we find that, independently of whether the stars
populate the hadronic or the twin branch, their response is characterised
by a critical-perturbation strength such that the star will oscillate on
the original branch for subcritical perturbations and migrate to the
neighbouring branch for supercritical perturbations while conserving
rest-mass. Because the critical values are different for stars with the
same rest-mass but sitting on either branch, it is possible to define as
favoured the part of the branch that has the largest critical
perturbation, thus correcting the common wisdom that stellar models on
the twin branch are the favoured ones. Interestingly, we show that the
binding energies on the two branches can be used to deduce without
simulations which of the stellar configurations is more likely to be
found in nature.
\end{abstract}

\maketitle

\section{Introduction}
\label{sec:introduction}

Neutron stars (NSs) gain pressing priority when it comes to understanding
the behaviour of matter at high density and low
temperatures~\cite{Rezzolla2018, baym_hadrons_2018,
  burgio_neutron_2021}. These astrophysical objects are harnessed by
immensely strong gravity, which allows them to compress the matter at the
core well beyond the nuclear saturation density, $\rho_\mathrm{sat}
\simeq 2.7 \times 10^{14} \, \mathrm{g/cm}^{3}$.  Quantum Chromodynamics
(QCD) predicts a phase transition (PT) from hadronic matter to quark
matter in such conditions, which could take place when a binary system of
neutron stars collides~\cite[see,
  \eg][]{paschalidis_implications_2018,Most:2018eaw,
  bauswein_identifying_2019, Weih:2019xvw, prakash_signatures_2021,
  huang_merger_2022, fujimoto_gravitational_2023, ujevic_reverse_2023,
  haque_effects_2024} or a supernova explosion takes place~\cite[see,
  \eg][]{sagert_signals_2009, zha_progenitor_2021,
  kuroda_core-collapse_2022, jakobus_role_2022}. This is a decades-old
hypothesis that is yet to meet its complete theoretical understanding
with complementing experimental (and observational) evidence.

The equation of state (EOS) of the cold neutron-rich matter is
sufficiently comprehended and mainly constrained by Chiral Effective
Field Theory~\cite[see, \eg][]{lonardoni_nuclear_2020,
  drischler_chiral_2021}. Relativistic heavy-ion collider experiments
\cite{adams_experimental_2005, arsene_quarkgluon_2005, back_phobos_2005,
  adcox_formation_2005} and their theoretical description~\cite[see,
  \eg][]{borsanyi_full_2014, nagata_finite-density_2022, Most2022e}
efficiently probe the matter at high temperatures and low densities, and
perturbative QCD calculations \cite{mogliacci_equation_2013,
  kurkela_cool_2016, alford_compact_2017,
  haque_next--next-leading-order_2021, christian_confirming_2022,
  gorda_equation_2023} give reliable results only at asymptotically high
densities. Hence, the NS cores, either when isolated or in binary
systems, have become the promising candidates to probe the
hadron-to-quark PT.

The onset of hadron-to-quark PT and the nature of this phenomenon are not
known a priori~\cite{oertel_equations_2017, blacker_constraining_2020,
  dore_far--equilibrium_2020}, which leaves a plethora of possible hybrid
EOSs that dictates the expression of PT inside the NS
core~\cite{annala_evidence_2020, verma_probing_2025}. Under these
assumptions, alternatives exist to the traditional NSs that are described
by purely hadronic EOSs. In particular, it is in principle possible to
consider the existence of ``hybrid stars'', that is, compact stars
comprising a quark-matter core and an hadronic-matter envelope.

In literature, PT has mainly been modelled as a transition using density
discontinuity at a constant pressure (Maxwell construction) or a
mixed-phase where hadrons and quarks coexist (Gibbs construction)
\cite{glendenning_phase_2001}. Importantly, the presence of a PT allows
for the existence of a new branch in the mass-radius sequence of stable
equilibria corresponding to hybrid stars. Two equilibria are then
  possible for the same mass on either of the stable branches. Hereafter,
  we will refer to as the ``twin branch'' (${\rm TB}$) the set of more
  compact hybrid stars stars, \ie the ``twin stars'', and as the
  ``hadronic branch'' (${\rm HB}$) the set of stellar models that are
  less compact for the same mass~\cite[see also,
  \eg][]{alford_generic_2013, christian_classifications_2018,
  montana_constraining_2019}.
  
The bulk of work produced over the last decade on twin stars is vast and
it is difficult to summarise it properly (the vast number of directions
pursued in the study of twins stars has been recently reviewed by
Ref.~\cite{haque_investigating_2026} and references [90, 105-159] therein
are particularly relevant for our results). Despite this rich literature,
the rather fundamental issue of the ``degeneracy of states'' in twin
stars has not been address yet. Stated differently, given the possible
existence of two equilibria for the same gravitational mass,
\textit{which of these two equilibria is the favoured one?} Since
equilibrium models are unable to break this degeneracy, we here address,
this simple and yet challenging question by studying for the first time
the response of twin and hadronic stars to nonlinear perturbations. Using
general-relativistic hydrodynamical simulations we establish under what
conditions a star on one of the two branches will ``migrate'' to the
neighbouring branch to seek a new and favoured equilibrium. Because these
migrations are not symmetric but reflect the different binding energies
of the equilibria on the two branches, we provide a simple and intuitive
criterion to determine which of the two solutions is the favoured
one. More in general, this study improves the long-standing question on
the nonlinear stability of equal-mass hydrostatic solutions of compact
stars in general relativity.

\section{Methods}
\label{sec:formalism}

\subsection{Equation of state}

A strong PT in EOS is identified by a region with vanishing speed of
sound, formed due to a jump in the energy density (or rest-mass density)
keeping the transition pressure constant
\cite{glendenning_phase_2001}. This region defines a Maxwell-type
first-order PT separating lower densities and higher densities given by
hadronic-matter and quark-matter EOSs, respectively. When performing
simulations in general-relativistic hydrodynamics, however, a vanishingly
small the speed of sound in the PT region needs special care to avoid
artefact in the numerical evolutions~\cite{font_three-dimensional_2002,
  dimmelmeier_dynamic_2009, Hanauske2018, espino_fate_2022,
  huang_phase-transition-induced_2025}. A simple solution to this problem
consists in constructing the EOS such that the pressure is not exactly
constant in the PT region and hence the speed of sound is very small but
nonzero.

As a generic representative of an EOS leading to a twin-star
configuration, we here consider the piecewise polytropic
prescription~\cite[see, \eg][]{Rezzolla_book:2013} adopted in Table 1
of~\cite{naseri_exploring_2024} where additional information can be
found. This study used the EOS for general-relativistic hydrodynamic
simulations of isolated hybrid stars, guaranteeing that the initial value
problem remains well-posed when the PT region is included in the stellar
profile. The six segments of the piecewise polytropic prescription
employed here are listed in Tab.~\ref{tab:ppeos}, where the crust is
described by the polytropes $i = 1, 2$, the hadronic part by the
polytropes $i = 3$, the PT region by $i = 4$, and, finally, the quark
region is described by the polytropes $i = 5, 6$ (note that because of
our simplified treatment of the crust, we use here only six of the eight
segments presented by~\cite{naseri_exploring_2024}).

\begin{table}
  \centering
  \begin{tabular}{@{\hspace{0.1cm}}c|c@{\hspace{0.2cm}}c@{\hspace{0.2cm}}c@{\hspace{0.2cm}}c@{\hspace{0.2cm}}c@{\hspace{0.2cm}}c@{\hspace{0.1cm}}}
  \hline
  $i$          & 1      & 2      & 3      & 4      & 5      & 6      \\
  \hline
  $\log k_i$   &  -4.245 & -21.197 & -35.544 & 10.108 & -64.283 & -101.225 \\
  $\Gamma_i$   & 1.139  & 2.354 & 3.346 & 0.258 & 5.188 & 7.610 \\
  $\log \rho_i$& 13.943 & 14.471 & 14.783 & 15.089 & 15.250 & --     \\
  \hline
  \end{tabular}
  \caption{Properties of the polytropic segments for the EOS used in this
    work in terms of the polytropic constants $k_i$, adiabatic indices
    $\Gamma_i$, and of the junction rest-mass densities $\rho_i$ in CGS
    units.}
  \label{tab:ppeos}
\end{table}

\begin{figure}
  \centering
  \includegraphics[width=0.45\textwidth]{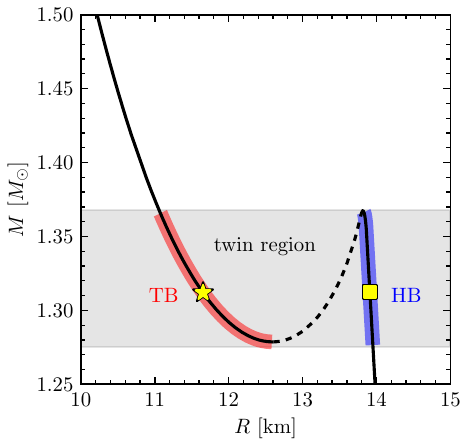}
  \caption{Representative example of a mass-radius sequence for an EOS
    leading to the coexistence of a hadronic branch (${\rm HB}$) and of a
    twin branch (${\rm TB}$). Solid and dashed lines are used to indicate
    branches that are stable or unstable to linear perturbations,
    respectively. Shown with a shaded area is the range in gravitational
    masses where the twin solutions exist, \ie the twin region. The
    yellow star on ${\rm TB}$ and the yellow box on ${\rm HB}$ indicate
    the models $\mathtt{TB.1.4}$ and $\mathtt{HB.1.4}$ which are selected
    to discuss the migration process in
    Sec.~\ref{sec:1.4Mb_evolution}. The properties of these
    configurations are listed in Tab.~\ref{tab:model}.}
  \label{fig:fig1}
\end{figure}

The sequences of stellar equilibrium models for the EOS considered is
reported in Fig.~\ref{fig:fig1} in terms of the gravitational mass $M$
and radius $R$, with the grey-shaded area the marking the ``twin
  region'', that is, the region in the mass-radius diagrams where the
hadronic branch (${\rm HB}$) and the twin branch (${\rm TB}$) coexist in
terms of rest-masses. Within the twin region, it is possible to use the
turning-point criterion $dM/d\rho_\mathrm{c}> 0$, where $\rho_\mathrm{c}$
is the central rest-mass density of the star to distinguish stable
stellar models (solid lines) from the unstable ones (dashed
lines);~\cite[see][for a discussion of the validity of this criterion
  for rotating and nonrotating stars]{Takami:2011}. We should note that a
very similar picture would be obtained if the gravitational mass $M$ in
Fig.~\ref{fig:fig1} were to be replaced by the rest-mass $M_{\rm
  b}$. This is because for these low-mass stars the two masses scale
essentially linearly~\cite[see, \eg Fig. 3 (left) of][]{Garibay2026}.

In both branches, we mark two stellar models having the same rest-mass
$M_\mathrm{b} = 1.4000\,M_{\odot}$ (yellow star for the ${\rm TB}$ and
yellow box for the ${\rm HB}$), which will be used to explain the
migration process. The properties of these stellar models are listed in
Tab.~\ref{tab:model}, where they are referred to as $\mathtt{TB.1.4}$ and
$\mathtt{HB.1.4}$, respectively. Note that despite having the same
rest-mass and essentially the same gravitational mass\footnote{We recall
that a quasi-universal relation can be used to relate $M$ and
${M}_\mathrm{b}$ analytically~\cite{Timmes1996, Garibay2026}.}, the two
configurations have radii that differ by $\gtrsim 2\, \mathrm{km}$,
making $\mathtt{TB.1.4}$ significantly more compact than
$\mathtt{HB.1.4}$. As a result, the matter in the core of
$\mathtt{TB.1.4}$ is compressed enough such that the central core is
described by a quark-matter EOS, while $\mathtt{HB.1.4}$ is purely made
of hadronic matter.

\begin{table}
  \centering
  \begin{tabular}{c@{\hspace{0.4cm}}c@{\hspace{0.4cm}}c@{\hspace{0.4cm}}c@{\hspace{0.4cm}}c}\\
    \hline
    Model & $\rho_c/\rho_\mathrm{sat}$ & $M~[M_\odot]$ & $M_{\mathrm{b}}~[M_\odot]$ & $R~[\mathrm{km}]$ \\
    \hline
    $\mathtt{TB.1.4}$ & 5.7342 & 1.3119 & 1.4000 & 11.6540 \\
    $\mathtt{HB.1.4}$ & 2.2227 & 1.3124 & 1.4000 & 13.9110 \\
    \hline
  \end{tabular}
  \caption{Properties of the representative stellar models indicated in
    Fig.~\ref{fig:fig1} with a yellow star and the yellow box. These
    models are selected to explore the migration process in
    Sec.~\ref{sec:1.4Mb_evolution}.}
  \label{tab:model}
\end{table}

\subsection{Numerical Setup}

To study the nonlinear response to perturbations of twin and hadronic
stars we have employed the open-source code \texttt{GR1D}
\cite{oconnor_new_2010}, which solves the spherically symmetric
general-relativistic hydrodynamic equations with a finite-volume scheme,
piecewise-parabolic reconstruction and an HLLE Riemann solver~\cite[see,
  \eg][]{Rezzolla_book:2013}.  The initial data is obtained via the
solution of the Tolman-Oppenheimer-Volkoff (TOV) equations.  A uniform
one-dimensional (1D) spatial grid is used with a resolution of $100\,{\rm
  m}$ and the outer boundary is extended to $200\,{\rm km}$ (additional
resolutions at $50\,{\rm m}$ and $25\,{\rm m}$ have been studied to
explore the converge of the solutions). The temporal evolution uses the
method-of-lines to a second-order Runge-Kutta scheme with a
Courant-Friedrichs-Lewy (CFL) condition set to 1/2. An atmosphere with a
constant density of $1\, \mathrm{g/cm}^3$ is set outside the stellar
profile. We finally note that because of the computational costs
associated with our large exploration of the space of parameters, which
counts more than 500 simulations, we have here resorted to the use of 1D
simulations.

\begin{figure*}
  \centering
  \includegraphics[width=0.8\textwidth]{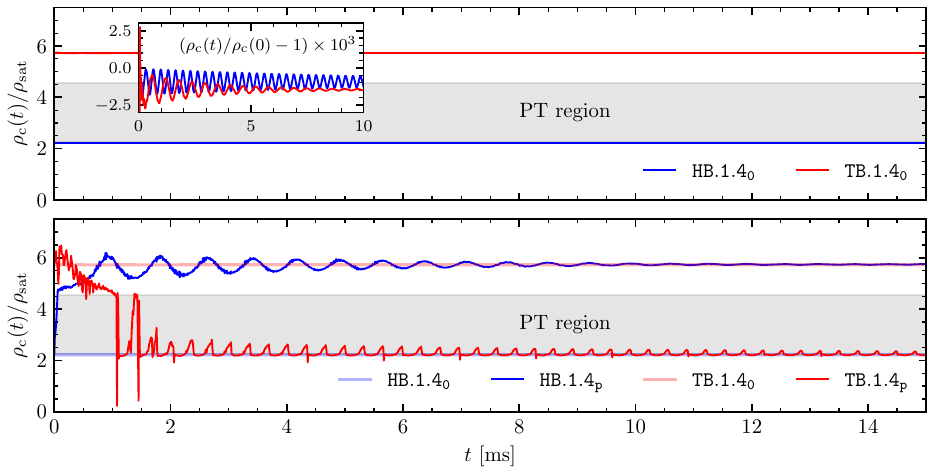}
  \caption{\textit{Top panel:} Evolution of the central rest-mass density
    $\rho_\mathrm{c}$ normalised to the nuclear-saturation density for
    the representative stellar configurations $\mathtt{TB.1.4}$ and
    $\mathtt{HB.1.4}$ that are shown with yellow symbols in
    Fig.~\ref{fig:fig1} (see also Tab.~\ref{tab:model}). Because no
    perturbations are introduced in this case, the inset is needed to
    show the minute relative oscillations and their decay with time. The
    shaded area reports the range of rest-mass densities across which the
    PT takes place. Clearly, both $\mathtt{TB.1.4}$ and $\mathtt{HB.1.4}$
    are stable to linear perturbations. \textit{Bottom panel:} the same
    as in the top but when the stars experience critical
    perturbations. In both cases, the stars migrate to the neighbouring
    branch of static solutions.}
  \label{fig:fig2}
\end{figure*}

\section{Results} 
\label{sec:results}

\subsection{Initial Perturbations}
\label{sec:init_pert}

As first shown in Ref.~\cite{font_three-dimensional_2002} using
three-dimensional general-relativistic simulations, a stellar model that
is in an unstable equilibrium can ``migrate'' to an equilibrium
configuration as a response to perturbations and while conserving
rest-mass. A similar process is expected to take place also in the case
of stars on the TB and HB, with the important difference that, in this
case, both of the stellar models are expected to be in a stable
equilibrium. Despite stability with respect to \textit{linear}
perturbations, one of the two stable configuration may be ``favoured''
over the other, so that, when subject to \textit{nonlinear}
perturbations, a star on the TB or HB will either oscillate in its
original branch or migrate to the neighbouring one. This migration can
either take place via an expansion to larger radii, \ie a ${\rm TB} \to
{\rm HB}$ transition, or via a compression to more compact
configurations, \ie a ${\rm HB} \to {\rm TB}$ transition.

To trigger such a transition, we introduce a perturbation that could
reflect a plausible astrophysical scenario, such as a stellar spin-down,
or the accretion of matter on the stellar surface, we introduce a
constant inward radial velocity of different strength on the initial
static stellar models, \ie
\begin{equation}
  \label{eq:pert_vr}
  v^r = 0 \qquad \to \qquad v^r =  - \lambda_{\rm H,T}\,,
\end{equation}
where $\lambda_{\rm H,T} > 0$ measures the strength of the perturbation
and the index ${\rm H,T}$ indicates whether the star is initially on the
${\rm HB}$ or on the ${\rm TB}$. Because extremely large perturbations
would be incompatible with the assumption of adiabatic evolutions, we
limit the strength of the initial perturbation to $\lambda_{\rm H,T} <
0.1\,c$; as we discuss below, such strength is sufficient to obtain all
the relevant results.

Ideally, it would be preferable to introduce perturbations with suitable
perturbative eigenfunctions in all hydrodynamical quantities. In
practice, and as done customarily in this type of studies~\cite[see,
  \eg][]{dimmelmeier_dynamic_2009, Radice2013c, shashank_f_2023,
  pierre_jacques_general_2025}, the solution of the full set of
hydrodynamical equations is such that over a couple of dynamical
timescales the star recovers the correct behaviour with the
eigenfunctions in the hydrodynamical variables matching well the
perturbative ones.

As mentioned above, the EOS employed here is meant to describe
zero-temperature nuclear matter in beta equilibrium. To reflect a purely
adiabatic treatment of the perturbations and concentrate on the simpler
interplay between kinetic energy and binding energy -- and hence not to
contaminate the dynamics with the inclusion of dissipative heating -- no
additional thermal contribution is added to the pressure during the
evolution, which is therefore isentropic. This is a reasonable
approximation given the modest velocity perturbations introduced.

\subsection{Migration dynamics}
\label{sec:1.4Mb_evolution}

The top panel of Fig.~\ref{fig:fig2} reports the evolutions of our two
reference stellar configurations without external perturbations, which we
will refer to as $\mathtt{HB.1.4_{0}}$ (black solid line) and
$\mathtt{TB.1.4_{0}}$ (red solid line), respectively. More specifically,
shown is the central rest-mass density $\rho_\mathrm{c}$ when normalised
to $\rho_\mathrm{sat}$. The two lines appear as perfectly horizontal here
but this is just the consequence of the scale used and the minuteness of
the perturbations that are triggered uniquely by the truncation error in
the discretisation of the equilibrium models. Shown however in the inset
is the central rest-mass density normalised to its initial value
$\rho_{\rm c}(t=0)$; the scale is sufficiently magnified so as to
appreciate that there indeed are oscillations and that these are of
$\simeq 0.1\%$. It is then clear that in both cases $\rho_\mathrm{c}$
oscillates around an equilibrium value\footnote{We note that the
equilibrium is towards slightly larger central rest-mass density values
because the numerical import naturally introduces a difference in the
equilibrium model~\cite[see][for a first discussion of this
  issue]{baiotti_three-dimensional_2005}.} and that the oscillations
eventually damp out at later time steps as a result of the small but
nonzero to numerical bulk viscosity~\cite{CerdaDuran2010, Chabanov2023b};
we have verified that both the amplitude and the decay rate of the
oscillations decrease as the resolution is increased, thus validating the
computational approach as numerically consistent
(see~\cite{Rezzolla_book:2013} and the discussion in
App.~\ref{app:res}). Also clear is that the oscillations have different
frequencies as expected for stars of different compactnesses, with
$\mathtt{HB.1.4_{0}}$ having an $f$-mode frequency that is $\sim1.74$
times larger than that of $\mathtt{TB.1.4_{0}}$. Overall, the results of
the top panel of Fig.~\ref{fig:fig2} indicate that both $\mathtt{HB.1.4}$
and $\mathtt{TB.1.4}$ are stable with respect to linear
perturbations. For a discussion of oscillation modes and stability
  of twin stars, see refs.~\cite{pereira_phase_2018,
    goncalves_fundamental-mode_2022, rau_two_2023, espino_fate_2022} and,
  for comparison with perturbative predictions in our case, see the
  discussion in App. \ref{app:fmode}.

By performing a very large number of simulations in which the strength of
the initial perturbation, \ie $\lambda_{\rm H,T}$, has revealed that a
threshold critical perturbation determines whether a perturbed star
(either on the ${\rm HB}$ or the ${\rm TB}$) will migrate to the
neighbouring branch. In other words, referring to Eq.~\eqref{eq:pert_vr}
there appears a critical value $\lambda_{\rm H, crit}$ such that a
perturbed star initially on the ${\rm HB}$ remain on the ${\rm HB}$ for
perturbations $\lambda_{\rm H} < \lambda_{\rm H, crit}$ while it will
migrate to the ${\rm TB}$ for $\lambda_{\rm H} \geq \lambda_{\rm H,
  crit}$. Similar considerations apply to stars initially on the ${\rm
  TB}$, such that the star will remain on the ${\rm TB}$ for
perturbations $\lambda_{\rm T} < \lambda_{\rm T, crit}$ while it will
migrate to the ${\rm HB}$ for $\lambda_{\rm T} \geq \lambda_{\rm T,
  crit}$.

To illustrate the dynamics of the migration, the bottom panel of
Fig.~\ref{fig:fig2} reports the evolution of the same models discussed in
the top panel but when a critical perturbation is used for the migration
on either side, \ie $\lambda_{\rm H} = \lambda_{\rm H, crit} = 0.021\,c$
and $\lambda_{\rm T} = \lambda_{\rm T, crit} = 0.040\,c$. We refer to
these models as $\mathtt{HB.1.4_{p}}$ and $\mathtt{TB.1.4_{p}}$,
respectively and report as a reference also in the bottom panel the
evolution of the models $\mathtt{HB.1.4_{0}}$ and $\mathtt{TB.1.4_{0}}$.

For the stellar model $\mathtt{HB.1.4_{p}}$, the initial infall velocity
perturbation induces a mini-collapse, with a rapid contraction of the
star and an increase in the central rest-mass density $\rho_\mathrm{c}$
over a timescale of $\sim 0.1\, \mathrm{ms}$. Since $\rho_\mathrm{c}$ of
$\mathtt{HB.1.4_{p}}$ (black solid line) is located close to the PT
region, the perturbation is strong enough to trigger a PT inside the core
and hence generate quark matter in the star. As a result, $\rho_\mathrm{c}$
continues to grow until the matter becomes sufficiently stiff enough to
exert an outward balancing pressure to halt the collapse around $t \sim
1\, \mathrm{ms}$. This results in $\rho_\mathrm{c}$ starting to oscillate
around an equilibrium value. As the evolution proceeds (we have carried
out the simulations up to $t \simeq 50\,{\rm ms}$),
the oscillations tend to decay and $\rho_\mathrm{c}$ settles around the
value corresponding to model $\mathtt{TB.1.4_{0}}$ with a precision of
$0.36\%$. After about $\sim 20\,{\rm ms}$, the system has reached its
new equilibrium and a successful ${\rm HB}\to {\rm TB}$ migration has
taken place.

A complementary dynamics is shown by the stellar model
$\mathtt{TB.1.4_{p}}$, where the initial radial kick similarly induces an
increase in $\rho_\mathrm{c}$, which however excites the high-frequency
oscillation modes. In this case, however, the collapse is immediately
halted by the stiff quark-matter EOS, thus resulting in a rapid bounce
and outward expansion of the whole star. The resulting decompression
allows for the central rest-mass density to enter at $t \sim 1\,{\rm ms}$
the high-density edge of the PT region in the EOS (the EOS at the PT
region can be thought to be extremely soft) and hence undergo a very
rapid expansion. As a result of the reverse PT, the star is purely
hadronic and the central density oscillates around the low-density edge
of the PT region while attaining a new equilibrium, which corresponds to
that of $\mathtt{HB.1.4_{0}}$ with a precision of $0.18\%$. Also in this
case, the oscillations are gradually damped and after about $\sim
20\,{\rm ms}$, the system has reached its new equilibrium and a
successful ${\rm TB}\to {\rm HB}$ migration has taken place.

\begin{figure}
  \centering
  \includegraphics[width=0.45\textwidth]{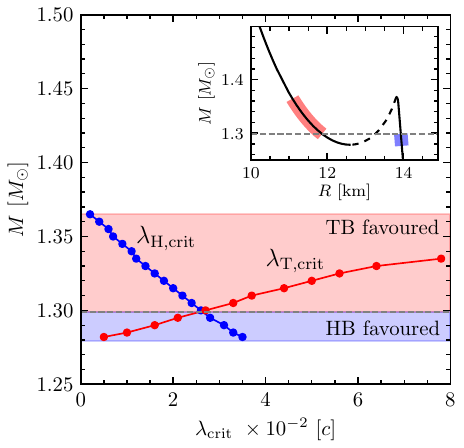}
  \caption{Numerical values of the critical perturbation velocities
    $\lambda_{\rm T, crit}$ for models on the ${\rm TB}$ (filled red
    circles) and the corresponding values $\lambda_{\rm H, crit}$ for
    models on the ${\rm HB}$ (filled blue circles) for the range of masses
    in the twin region. The different colours in the two shaded regions
    show which of the two branches is favoured when comparing
    $\lambda_{\rm T, crit}$ and $\lambda_{\rm H, crit}$; shown with a
    horizontal dashed line is the value of the neutral-favouritism mass
    $M_\mathrm{neut} = 1.2988 \, M_\odot$, where the two branches are
    equally favoured. Shown in the inset is a representation of the
    favourite branches in an $(M,R)$ diagram.}
\label{fig:fig3}    
\end{figure}

\subsection{Critical perturbations and favoured twin solutions}

The simulations described above for the $\mathtt{HB.1.4}$ and
$\mathtt{TB.1.4}$ models have been repeated for a large number of stellar
models with different baryonic (and gravitational) masses where,
$\lambda_\mathrm{crit}$ was adaptively computed using a bisection process
up to the accuracy of $0.001\,c$. The dynamics is then essentially the
same but with the important difference that the critical perturbation
velocity is a function of the mass, \ie $\lambda_{\rm T, crit} =
\lambda_{\rm T, crit}(M)$ and $\lambda_{\rm H, crit} = \lambda_{\rm H,
  crit}(M)$. Figure~\ref{fig:fig3} highlights this mass dependence by
reporting the critical values for both branches as a function of the
gravitational mass. Interestingly, the behaviour of the two critical
perturbations with $M$ is very different, so that, in general,
$\lambda_{\rm T, crit} \neq \lambda_{\rm H, crit}$.

These results are very revealing as they show that although migrations in
both directions are possible, \ie ${\rm HB} \leftrightarrows {\rm TB}$,
they are not equally likely. On the contrary, a star on a given branch
will prefer to stay on that branch unless it is subject to a very large
(and possibly unrealistic) perturbation is experienced. In turn, this
allows us to introduce criterion discriminating which of the two twins is
the ``favoured'' one. More specifically, we conclude that 
\begin{align}
  \label{eq:_TB_fav}
  &{\rm TB~~is~~favoured~~if} && \lambda_{\rm T, crit}(M) <
  \lambda_{\rm H, crit}(M)\,,& \\
  \label{eq:_HB_fav}
  &{\rm HB~~is~~favoured~~if} && \lambda_{\rm T, crit}(M) >
  \lambda_{\rm H, crit}(M)\,,&
\end{align}
with a ``neutral favouritism'' condition (analogous to a ``neutral
stability'' condition) when
\begin{equation}
\lambda_{\rm T, crit} (M=M_{\rm neut}) =
\lambda_{\rm H, crit} (M=M_{\rm neut})\,.
\end{equation}
For the specific EOS considered here, this latter condition is met for a
neutral mass $M_{\rm neut}=1.2988 \, M_\odot$ and is marked by an
horizontal dashed line in Fig.~\ref{fig:fig3}. It may be useful to remark
that while the precise values of $\lambda_{\rm T, crit}$ and
$\lambda_{\rm H, crit}$ will depend (weakly) on the type of perturbations
chosen (see App.~\ref{app:positive_pert}), the definitions
\eqref{eq:_TB_fav} and \eqref{eq:_HB_fav} will continue to hold. More
Importantly, the neutral mass $M_{\rm neut}$ will be independent of the
perturbation chosen and thus always the same for a given EOS.

Overall, the response of twin and hadronic stars to perturbations nicely
summarised in Fig.~\ref{fig:fig3} points out what is possibly the most
important result of this paper, namely, that the prevailing notion that
stellar models on the TB are those more likely to be found in nature is
incomplete and rather conditional on the position of $M_\mathrm{crit}$ in
the twin region. The response to perturbations analysed here clearly
shows that low-mass hadronic stars are favoured over the corresponding
twin solution. Furthermore, as we will discuss in
Sec.~\ref{sec:large_PTs}, the actual existence of stars on the TB may is
highly unlikely if the PT is very large.

\begin{figure}
  \centering
  \includegraphics[width=0.45\textwidth]{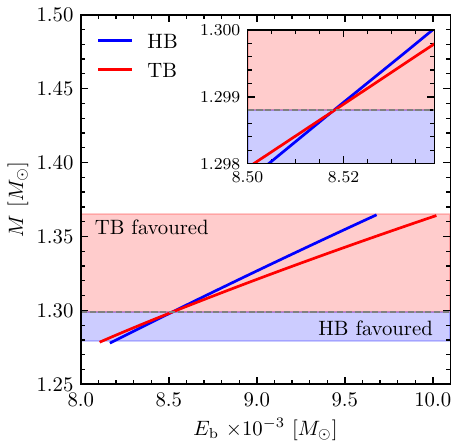}
  \caption{The same as in Fig.~\ref{fig:fig3} but in terms of the binding
    energy $E_\mathrm{b}:= M_{\rm b} - M$; the inset shows a magnified
    view of the binding energies near the neutral mass $M_{\rm
      neut}=1.2988 \, M_\odot$. Clearly, the representations in terms of
    $E_\mathrm{b}$ and $\lambda_\mathrm{crit}$ are very similar, so that
    the binding energies can be used to deduce the favoured twin without
    resorting to numerical simulations.}
\label{fig:fig4}    
\end{figure}

The existence of favoured and neutral stellar configurations can be
intuitively understood if one bears in mind that stable equilibrium
stellar models can be associated in a classical sense to local ground
states of the Einstein equations, therefore as sitting at the bottom of
the local minimum of an effective potential that depends on the
perturbation strength. When subject to small (linear) perturbations,
their energy will be varied by the introduction of the energy in the
perturbation, which will induce oscillations around the ground states. So
long as the energy perturbation is smaller than the potential barrier,
the configuration will remain in the same local (and false) minimum or,
equivalently, a star in a favoured branch will remain in such a
branch. However, if the energy in the perturbation is sufficiently large
to overcome the potential barrier, a second local minimum will become the
true minimum and the star will move to it. This is what happens when a
star on a given branch migrates to the neighbouring one for $\lambda_{\rm
  H,T} \geq (\lambda_{\rm H,T})_{\rm crit}$. Clearly, if the two minima
are identical, both branches are equally favoured, hence the existence of
a neutral mass where this happens. The concept of neutral mass and the
distinction between the favoured branches in the $(M,R)$ plane is
summarised in the inset of Fig.~\ref{fig:fig3}.

If the existence of a neutral mass $M_{\rm neut}$ is rather natural, a
final point to clarify is what determines it. This can be easily
addressed by comparing the gravitational binding energies $E_\mathrm{b}
:= M_\mathrm{b} - M > 0$ of the stellar models on the two branches. This
is shown in Fig.~\ref{fig:fig4}, which reports the binding energy for
stars on the ${\rm HB}$ (blue solid line) and on the ${\rm TB}$ (red
solid line). Although these energies are very similar, they are not
identical and, interestingly, they coincide at $M = M_{\rm neut}$. The
mapping between $\lambda_{\rm crit}$ and $E_{\rm b}$ is true with a
precision of $0.1\%$. This result clearly illustrates that, for any
gravitational mass, the favoured stars are those with the largest binding
energy as they are sitting in the true equilibrium state. Furthermore, it
shows that the potential-barrier differences -- which are effectively
mapped to the difference $|\lambda_{\rm H, crit} - \lambda_{\rm T,
  crit}|$ -- are smaller for $M < M_{\rm neut}$, which also explains why
in Fig.~\ref{fig:fig3} the difference in critical migration values, \ie
$|\lambda_{\rm H, crit} - \lambda_{\rm T, crit}|$, is smaller for masses
below the neutral one.

\subsection{Other classes of twin stars}
\label{sec:large_PTs}

\begin{figure}
  \centering
  \includegraphics[scale=0.66]{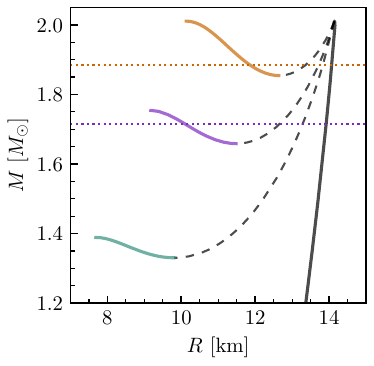}
  \includegraphics[scale=0.66]{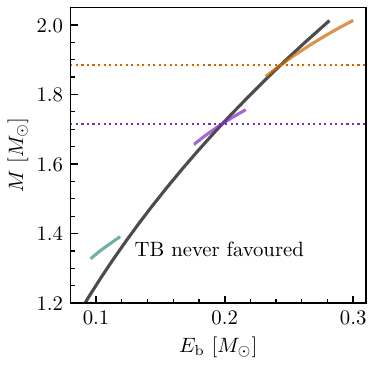}
  \caption{\textit{Left panel:} the same as in Fig.~\ref{fig:fig1} but
    for three EOSs that lead to twin stars of category II. The stable
    ${\rm HB}$ is marked with a black solid line, while the stable ${\rm
      TBs}$ are marked with solid lines of different colours; dashed and
    dotted lines mark the (linearly) unstable branches and the neutral
    masses $M_{\rm neut}$, respectively. \textit{Right panel:} the same
    as in Fig.~\ref{fig:fig4} but for the three EOSs on the left panel;
    also here, the coloured dotted lines mark the neutral masses. Note
    that no neutral mass exists for the EOS with the smallest maximum
    mass on the ${\rm TB}$; hence, no stellar model on the ${\rm TB}$ is
    favoured for this EOS. Prescriptions of these EOSs are discussed in
    App.~\ref{app:eos}.}
    \label{fig:fig5}    
\end{figure}

The picture summarised above refers to twin stars that are classified to
belong to category III~\cite{christian_classifications_2018,
  montana_constraining_2019}, that is, stars for which the maximum mass
of the ${\rm HB}$ is smaller than the maximum mass of the ${\rm TB}$, \ie
$M_{\mathrm{TOV, HB}} < M_{\mathrm{TOV, TB}}$; for obvious reasons, a
very similar response is expected also for categories I, and IV of the
same classification. However, there is a fourth category, for which the
opposite is true, that is, $M_{\mathrm{TOV, TB}} \leq M_{\mathrm{TOV,
    HB}}$, and this is referred to as category
II~\cite{christian_classifications_2018,
  montana_constraining_2019}. These configurations are shown in the left
panel of Fig.~\ref{fig:fig5}, which reports three different EOSs with
$M_{\mathrm{TOV, TB}} = M_{\mathrm{TOV, HB}} = 2.010\,M_{\odot}$, but
also $M_{\mathrm{TOV, TB}} = 1.754\,M_{\odot}$ and $1.389\,M_{\odot}$.

The response of these compact-star configurations to perturbations can be
easily deduced without performing numerical simulations (which we have
however performed), by simply looking at the binding energies of the
different stellar models. These are reported in the right panel of
Fig.~\ref{fig:fig5} and it then appears clear that while the first two
EOSs from the top lead to binding-energy curves that cross at some
specific value of the neutral mass (\ie $M_{\rm neut} = 1.886\,M_{\odot}$
and $1.716\,M_{\odot}$) the third EOS leads to sequences of stellar
models whose binding energies never intersect. As a result, for this EOS,
no neutral mass exists and only one branch will be favoured. Indeed, this
is exactly what we find via numerical simulations, which reveal that
$\lambda_{\rm T,crit} < \lambda_{\rm H,crit}$ for all stars on the TB
with $M \lesssim 1.355\,M_{\odot}$.  At the same time, the response to
perturbations of stars on the TB having mass $M \gtrsim 1.355\,M_{\odot}$
is such that they will not migrate towards the HB, but rather collapse to
black holes as a result of the radial-velocity
perturbation. Unsurprisingly, smaller and smaller perturbations are
needed as the mass of the perturbed star approaches $M_{\mathrm{TOV,
    TB}}$. Overall, this phenomenology reveals a very intriguing new
behaviour of very compact twin-star configurations, where a new
criticality in the behaviour -- \ie the criticality to collapse to black
holes -- appears in addition to criticality to migration. We will explore
this in more detail in a forthcoming paper.

\section{Conclusion}
\label{sec:conclusion}

Twin stars are the manifestation of the occurrence of a first-order PT
and have been the subject of numerous investigations over the last
decade. While it is simple to distinguish which of the compact-star
solutions resides on a branch that is stable to linear perturbations,
much more difficult is to predict which of the twin solutions is the
``favoured'' one, that is, which of the two possible stable
configurations is expected to be found in nature. We have addressed this
simple but fundamental question by studying via fully
general-relativistic simulations the response of stars on the TB and HB
to nonlinear perturbations triggered by an initial inward radial velocity
of various strength.

In this way, we have found that the response of all stellar models,
independently of whether they sit on the hadronic or twin branch, is
characterised by a critical-perturbation strength. For subcritical
perturbations the star will oscillate on the original branch with damped
oscillations, while for supercritical perturbations it will migrate to
the neighbouring branch, where it will eventually settle to the
corresponding equilibrium having the same rest-mass. More importantly,
the critical values on the hadronic and twin branches depend on the mass
of the star but are different for the same mass, so that it is possible
to define as ``favoured'' the part of the branch which has the largest
critical perturbation, as one expects that this represents the true
ground system of the space of configurations.

Interestingly, we have also found that the differences in critical
perturbations can be mapped into differences in the binding energies of
the stellar models on the two branches. As a result, the favoured star is
simply that with the largest binding energy and a (neutral) mass appears
where the favouritism is neutral and both configurations are equally
favoured. Hence, a simple investigation of the binding energies on the
two branches will reveal which of the stellar configurations is more
likely to be found in nature. Finally, while our results have been
applied to an EOS leading to a specific class of twin stars, \ie
categories I, III, and IV of the classification
in~\cite{christian_classifications_2018, montana_constraining_2019}, we
have shown, both with simulations and with the binding energies, that the
conclusions drawn above hold true also in other classes of twin stars.

Overall, these results demonstrate that the common wisdom for which the
stellar models on the twin branch are the one expected to revealed by
observations is, at least in part, incorrect. Indeed, it is reasonable to
expect stars that should populate the low-mass part of the TB are not
found in nature, and that the corresponding models should exist as stable
hadronic stars or black holes.

Finally, while this study significantly improves our understanding of the
long-standing issue of the existence of twin-star solutions, it can be
extended in a number of ways. First, by a more extensive validation of
the location of the critical perturbations via 3D simulations and the
inclusion of non-spherical perturbations. Second, by the inclusion of a
broader set of EOSs that may reveal quasi-universal properties in the
neutral-favouritism mass $M_{\rm neut}$, which, in turn, could be used to
set tighter constraints on the existence of twin stars. Third, by
extending the set of initial configurations considered to include
configurations that are rotating and magnetised. We plan to explore these
aspects in future work.

\begin{acknowledgments}

It is a pleasure to thank Mark Alford, M. Hanauske, J.
Schaffner-Bielich, A. Sedrakian, S. Chatterjee, A. Karan, T. P. Jacques,
and Samuel Tootle for helpful discussions and useful feedback. SH and RM
acknowledge the SERB grant: CRG/2022/000663. Partial funding comes from
the ERC Advanced Grant ``JETSET: Launching, propagation and emission of
relativistic jets from binary mergers and across mass scales'' (Grant
No. 884631). LR acknowledges the Walter Greiner Gesellschaft zur
F\"orderung der physikalischen Grundlagenforschung e.V. through the Carl
W. Fueck Laureatus Chair. The following packages has been used in this
work---{\sc GR1D}~\cite{oconnor_new_2010}, {\sc
  Numpy}~\cite{harris_array_2020}, {\sc
  Matplotlib}~\cite{hunter_matplotlib_2007}, {\sc
  Jupyter}~\cite{kluyver_jupyter_2016}.
\end{acknowledgments}

\appendix

\begin{table*}
  \centering
  \begin{small}
  \begin{tabular}{@{\hspace{0.1cm}}l|r@{\hspace{0.15cm}}r@{\hspace{0.15cm}}r@{\hspace{0.15cm}}r@{\hspace{0.15cm}}r@{\hspace{0.15cm}}r@{\hspace{0.15cm}}r@{\hspace{0.15cm}}r@{\hspace{0.15cm}}r@{\hspace{0.15cm}}r@{\hspace{0.15cm}}r@{\hspace{0.15cm}}r@{\hspace{0.15cm}}r@{\hspace{0.15cm}}r@{\hspace{0.15cm}}r@{\hspace{0.1cm}}}
  \hline
  $i$   & 1~~~~& 2~~~~& 3~~~~& 4~~~~& 5~~~~& 6~~~~& 7~~~~& 8~~~~& 9~~~~& 10~~~~& 11~~~~& 12~~~~& 13~~~~& 14~~~~& 15~~~~ \\
  \hline
  $\log k_i$  & -8.167 & -5.974 & 1.725 & -7.399 & -24.415 & -61.005 & 11.099 & -64.768 & -53.435 &  -45.867 & -39.803 & -35.246 & -31.443 &  -28.394 & -25.796 \\ 
  $\Gamma_i$  & 1.584  & 1.287 & 0.622 & 1.356 & 2.570 & 5.100  & 0.200  & 5.250 & 4.500 & 4.000 & 3.600 & 3.300 & 3.050 & 2.850 & 2.680 \\
  $\log \rho_i$   & 7.385 & 11.577 & 12.419 & 14.027 & 14.462 & 14.715 & 15.023 & 15.110 & 15.136 & 15.159 & 15.187 & 15.214 & 15.245 & 15.278 & --- \\
  \hline
  $\log k_i$  & -8.167 & -5.974 & 1.725 & -7.399 & -24.415 & -61.005 & 11.099 & -65.226 & -53.810 & -46.187 & -40.079 & -35.490 & -31.658 & -28.586 & -25.970 \\ 
  $\Gamma_i$  & 1.584  & 1.287 & 0.622 & 1.356 & 2.570 & 5.100  & 0.200  & 5.250 & 4.500 & 4.000 & 3.600 & 3.300 & 3.050 & 2.850 & 2.680 \\
  $\log \rho_i$   & 7.385 & 11.577 & 12.419 & 14.027 & 14.462 & 14.715 &  15.113 & 15.220 & 15.245 & 15.270 & 15.298 & 15.326 & 15.357 & 15.389 & --- \\
  \hline
  $\log k_i$  & -8.167 & -5.974 & 1.725 & -7.399 & -24.415 & -61.005 & 11.099 & -66.000 & -54.446 & -46.730 & -40.548 & -35.903 & -32.025 & -28.917 & -26.269 \\ 
  $\Gamma_i$  & 1.584  & 1.287 & 0.622 & 1.356 & 2.570 & 5.100  & 0.200  & 5.250 & 4.500 & 4.000 & 3.600 & 3.300 & 3.050 & 2.850 & 2.680 \\
  $\log \rho_i$   & 7.385 & 11.577 & 12.419 & 14.027 & 14.462 & 14.715 &  15.267 & 15.405 & 15.431 & 15.455 & 15.482 & 15.511 & 15.541 & 15.573 & --- \\
  \hline
  \end{tabular}
  \end{small}
  \caption{Properties of the polytropic segments for the EOSs used in
    Sec.~\ref{sec:large_PTs} in terms of the polytropic constants $k_i$,
    adiabatic indices $\Gamma_i$, and of the junction rest-mass densities
    $\rho_i$ in CGS units. The $M$--$R$ sequences are plotted in the left
    panel of Fig.~\ref{fig:fig5}. The top, middle, and bottom sets of
    values form EOSs that construct orange, purple, and green hybrid
    branches, respectively, with a common hadronic branch.}
  \label{tab:ppeos_more}
\end{table*}

\section{Supplementary Information}
\label{app:conv}

\subsection{Resolution dependence}
\label{app:res}

\begin{figure}
  \centering
  \includegraphics[width=0.47\columnwidth]{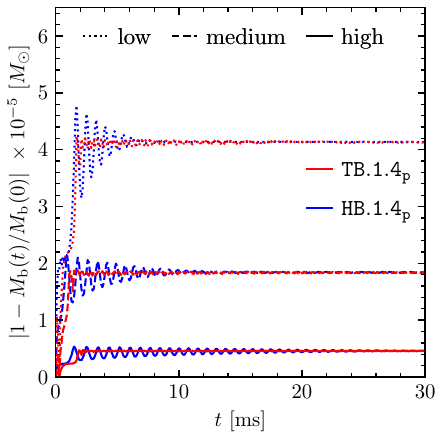}
  \hskip 0.1cm
  \includegraphics[width=0.47\columnwidth]{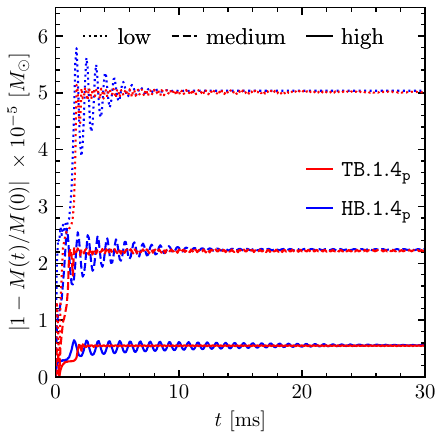}
  \caption{Evolution of the relative differences in the rest-mass $M_{\rm
      b}$ (left panel) and of the gravitational mass $M$ (right panel)
    when compared to the initial values, \ie $|1 - M_{\rm b}(t)/M_{\rm
      b}(0)|$ (left panel) and $|1 - M(t)/M(0)|$ (right panel),
    respectively. Different lines show the variations for the three
    different resolutions considered here, namely, $\Delta
    x=150\,\mathrm{m}$ (low; dotted line), $\Delta x=100\,\mathrm{m}$
    (medium; dashed line), and $\Delta x=50\,\mathrm{m}$ (high; solid
    line); the additional ``very-high'' resolution with $\Delta
    x=25\,\mathrm{m}$ was computed but it is not reported as it would not
    be visible on these scales. Different colours refer to either the
    perturbed model $\mathtt{TB.1.4_{p}}$ (red lines) or to perturbed
    model $\mathtt{HB.1.4_{p}}$ (blue lines).}
    \label{fig:fig_m}    
\end{figure}

We first discuss the robustness of our conclusion and present evidence
that the resolution employed is adequate to measure the critical
perturbation velocity inducing the migration. Taking as reference models
$\mathtt{TB.1.4}$ and $\mathtt{HB.1.4}$ (see Tab.~\ref{tab:model}), we
recall that they have binding energies $E_\mathrm{b}=0.0881\,M_\odot$ and
$E_\mathrm{b}=0.0876\,M_\odot$, respectively. Since the differences in
these two binding energies is $\sim 10^{-4}\,M_\odot$, the evolutions
must have a precision that is smaller (or much smaller) than this
difference in order to capture robustly the migration process. In
Fig.~\ref{fig:fig_m}, we report the evolution of the relative differences
in the rest-mass $M_{\rm b}$ and of the gravitational mass $M$ when
compared to the initial values, \ie $|1 - M_{\rm b}(t)/M_{\rm b}(0)|$
(left panel) and $|1 - M(t)/M(0)|$ (right panel)\footnote{We recall that
the gravitational mass is numerically extracted at the outer boundary
after a volume integral across the whole domain [see Eq.(4) in
  \cite{oconnor_open-source_2015}].}, respectively.  Note that the data
refers to the perturbed models, \ie $\mathtt{HB.1.4_{p}}$ (blue lines)
and $\mathtt{TB.1.4_{p}}$ (red lines), since the unperturbed ones
($\mathtt{HB.1.4_{0}}$, $\mathtt{TB.1.4_{0}}$) have even smaller
variations, which are about two orders of magnitude smaller.

Different lines show the variations for the three different resolutions
considered here, namely $\Delta x=150\,\mathrm{m}$ (low), $\Delta
x=100\,\mathrm{m}$ (medium and default resolution), $\Delta
x=50\,\mathrm{m}$ (high), and $\Delta x=25\,\mathrm{m}$ (very high).
Clearly, the figure highlights that even when employing the low
resolution, the relative variations in the rest mass with and in the
gravitational mass are of the order of $10^{-5}$ and that this these
reduce to values $\sim 10^{-6}$ when the resolution is
increased. Similarly, Tab.~\ref{tab:conv} reports the values of the
critical velocity perturbation $\lambda_{\rm crit}$ and highlights that
the values computed converge to a value that is constant with a precision
of $0.1\%$. Overall, the data in Fig.~\ref{fig:fig_m} and
Tab.~\ref{tab:conv} demonstrates that the resolutions employed are more
than sufficient to capture the nonlinear developments in the evolution.

\begin{table}
  \centering
  \begin{tabular}{@{\hspace{0.1cm}}c@{\hspace{0.3cm}}@{\hspace{0.3cm}}|c@{\hspace{0.3cm}}|c@{\hspace{0.3cm}}|c@{\hspace{0.3cm}}|c@{\hspace{0.1cm}}}
    \multicolumn{5}{c}{$\lambda_\mathrm{crit}~[{c}]$} \\
    \hline
    Model & ${\rm low}$ & ${\rm medium}$ & ${\rm high}$ & ${\rm very~high}$ \\
    & $(150\,\mathrm{m})$ & $(100\,\mathrm{m})$ & $(50\,\mathrm{m})$ & $(25\,\mathrm{m})$ \\
    \hline
    $\mathtt{TB.1.4}$ & 0.04790 & 0.03963 & 0.03547 & 0.03415 \\
    $\mathtt{HB.1.4}$ & 0.02040 & 0.02028 & 0.02022 & 0.02019 \\
    \hline
  \end{tabular}
  \caption{Variation of $\lambda_\mathrm{crit}$ for three different
    resolutions from low to high and for the representative stellar
    models given in Tab.~\ref{tab:model}. Note how the values converge to
    a constant as the resolution is increased.}
  \label{tab:conv}
\end{table}

\subsection{Variations on the perturbation profiles}
\label{app:positive_pert}

As mentioned in the main text, while the value of the neutral mass
$M_{\rm neut}$ is independent of the perturbation chosen and thus always
the same for a given EOS, the precise values of $\lambda_{\rm T, crit}$
and $\lambda_{\rm H, crit}$ will depend on the type of perturbations
chosen. To validate this dependence, we have considered the migration
dynamics using velocity perturbation profiles with different
profiles. More specifically, we adopt a two-parameter perturbation
velocity function proposed by~\cite{naseri_exploring_2024},
\begin{equation}
  \label{eq:vr_generic}
  v^r = -\Tilde{\lambda}\left(\dfrac{r}{R}\right)^\alpha\,,
\end{equation}
where $\Tilde{\lambda}$ decides the perturbation strength and the
exponent $\alpha$ controls the radial profile of the perturbation. For
$\alpha=0$, the function reduces to the form of Eq.~\eqref{eq:pert_vr},
\ie the perturbation profile used for our study. For $\alpha>0$, the
perturbation smoothly goes to zero at the centre.

In Tab.~\ref{tab:more_pert}, we summarise the values of
$\Tilde{\lambda}_\mathrm{crit}$ for a variety of perturbation
profiles. Note how the values of the critical depend only weakly on the
profile chosen, but also that larger perturbations are needed for larger
values of the exponent $\alpha$. This has a simple justification: an
increase in the exponent $\alpha$ decreases the perturbation strength
inside the star, so that the linearly stable stars will require stronger
perturbation strengths to migrate. 

\begin{table}
  \centering
  \begin{tabular}{@{\hspace{0.1cm}}c@{\hspace{0.3cm}}|@{\hspace{0.3cm}}c@{\hspace{0.5cm}}c@{\hspace{0.5cm}}c@{\hspace{0.5cm}}c@{\hspace{0.1cm}}}
     \multicolumn{5}{c}{$\Tilde{\lambda}_\mathrm{crit}~[{c}]$} \\
    \hline
    Model & $\alpha=0$ & $\alpha=0.5$ & $\alpha=1.0$ & $\alpha=2.0$ \\
    \hline
    $\mathtt{TB.1.4}$ & 0.040 & 0.049 & 0.060 & 0.089 \\
    $\mathtt{HB.1.4}$ & 0.021 & 0.030 & 0.040 & 0.066 \\
    \hline
  \end{tabular}
  \caption{Comparison of $\Tilde{\lambda}_\mathrm{crit}$ for different
    perturbation profiles given by Eq.~\eqref{eq:vr_generic} for the
    representative stellar models given in Tab.~\ref{tab:model}.}
  \label{tab:more_pert}
\end{table}

\subsection{Comparison of stable evolution with perturbative results} 
\label{app:fmode}

As anticipated in the main text, the stars on TB and HB are
  linearly stable and hence will undergo (damped) oscillations when
  exposed to linear perturbations. The eigenfrequencies and
  eigenfunctions of such oscillations can be computed via linear
  perturbation theory or via nonlinear numerical simulations so long as
  the perturbations are very small and the response can be considered to
  be in a linear regime. Obviously, the two approaches can be compared
  and must yield, for instance, the same eigenfrequencies~(see, e.g.,
  Ref.~\cite{font_three-dimensional_2002} for a first systematic
  comparison in 3D simulations).

\begin{figure}
  \centering
  \includegraphics[width=0.45\textwidth]{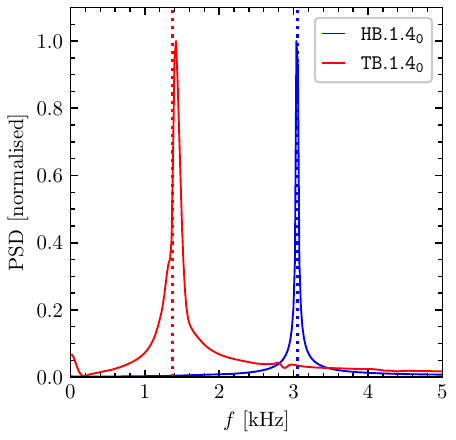}
  \caption{Power spectral density (PSD) of the numerical evolution
      of central rest-mass density $\rho_c(t)$ for models
      $\mathtt{HB.1.4_{0}}$ (blue solid line) and $\mathtt{TB.1.4_{0}}$
      (red solid line) after an evolution of $50\, \mathrm{ms}$. Reported
      with vertical dotted lines of the same colour are the corresponding
      eigenfrequencies of the $f$-mode as computed from perturbation
      theory.}
  \label{fig:fig_fmode}
\end{figure}

Here, we compare the numerical $f$-mode eigenfrequency as extracted
  from the numerical oscillations of the central rest-mass density
  $\rho_c(t)$ for the models $\mathtt{HB.1.4}_\mathtt{0}$ and
  $\mathtt{TB.1.4}_\mathtt{0}$ after a perturbation triggered by the
  truncation error only, with the radial $f$-mode computed from
  perturbative approach \cite{kumar_modification_2025, karan_private}.
  Figure~\ref{fig:fig_fmode} reports the power spectral density of the
  numerical evolution of $\rho_c(t)$ for models $\mathtt{HB.1.4_{0}}$
  (blue solid line) and $\mathtt{TB.1.4_{0}}$ (red solid line) after an
  evolution of $50\, \mathrm{ms}$. Also reported with vertical dotted
  lines of the same colour are the corresponding eigenfrequencies of the
  $f$-mode as computed from perturbation theory. Clearly, the match is
  very good, with only a very small difference ($\lesssim 4\%$), which is
  however inevitable as the numerical solution is always nonlinear even
  for very small perturbations. Overall, this result demonstrate the
  ability of our numerical simulations to explore accurately both the
  linear and the nonlinear regimes of our stellar models.

\subsection{Variations on the equation of state} 
\label{app:eos}

As discussed in Sec.~\ref{sec:large_PTs}, we have also considered the
evolution of other classes of twin stars. The fifteen segments of the
piecewise polytropic prescription employed here are listed in
Tab.~\ref{tab:ppeos_more}. A common crust part, implemented
from~\cite{read_constraints_2009}, is described by the polytropes $i =
1$--$4$, and a common hadronic part is described by the polytropes $i =
5,6$, making the EOSs identical up to $\rho_6$. This results in a common
hadronic branch, as indicated by black curve in the left panel of
Fig.~\ref{fig:fig5}. The PT region is described by the polytropic piece
with $i = 7$, in which $\Delta \rho = \rho_7 - \rho_6$ dictates the
strength of PT. We set $\Delta \rho_\mathrm{orange} < \Delta
\rho_\mathrm{purple} < \Delta \rho_\mathrm{green}$, which leads to a more
compact green hybrid branch with a smaller value for
$M_\mathrm{TOV,TB}$. Finally, the quark region is described by the
polytropes $i = 8$--$15$.

\bibliography{main}

\providecommand{\noopsort}[1]{}
\begin{thebibliography}{71}%
\makeatletter
\providecommand \@ifxundefined [1]{%
 \@ifx{#1\undefined}
}%
\providecommand \@ifnum [1]{%
 \ifnum #1\expandafter \@firstoftwo
 \else \expandafter \@secondoftwo
 \fi
}%
\providecommand \@ifx [1]{%
 \ifx #1\expandafter \@firstoftwo
 \else \expandafter \@secondoftwo
 \fi
}%
\providecommand \natexlab [1]{#1}%
\providecommand \enquote  [1]{``#1''}%
\providecommand \bibnamefont  [1]{#1}%
\providecommand \bibfnamefont [1]{#1}%
\providecommand \citenamefont [1]{#1}%
\providecommand \href@noop [0]{\@secondoftwo}%
\providecommand \href [0]{\begingroup \@sanitize@url \@href}%
\providecommand \@href[1]{\@@startlink{#1}\@@href}%
\providecommand \@@href[1]{\endgroup#1\@@endlink}%
\providecommand \@sanitize@url [0]{\catcode `\\12\catcode `\$12\catcode `\&12\catcode `\#12\catcode `\^12\catcode `\_12\catcode `\%12\relax}%
\providecommand \@@startlink[1]{}%
\providecommand \@@endlink[0]{}%
\providecommand \url  [0]{\begingroup\@sanitize@url \@url }%
\providecommand \@url [1]{\endgroup\@href {#1}{\urlprefix }}%
\providecommand \urlprefix  [0]{URL }%
\providecommand \Eprint [0]{\href }%
\providecommand \doibase [0]{https://doi.org/}%
\providecommand \selectlanguage [0]{\@gobble}%
\providecommand \bibinfo  [0]{\@secondoftwo}%
\providecommand \bibfield  [0]{\@secondoftwo}%
\providecommand \translation [1]{[#1]}%
\providecommand \BibitemOpen [0]{}%
\providecommand \bibitemStop [0]{}%
\providecommand \bibitemNoStop [0]{.\EOS\space}%
\providecommand \EOS [0]{\spacefactor3000\relax}%
\providecommand \BibitemShut  [1]{\csname bibitem#1\endcsname}%
\let\auto@bib@innerbib\@empty
\bibitem [{\citenamefont {Rezzolla}\ \emph {et~al.}(2018)\citenamefont {Rezzolla}, \citenamefont {Pizzochero}, \citenamefont {Jones}, \citenamefont {Rea},\ and\ \citenamefont {Vida{\~n}a}}]{Rezzolla2018}%
  \BibitemOpen
  \bibinfo {editor} {\bibfnamefont {L.}~\bibnamefont {Rezzolla}}, \bibinfo {editor} {\bibfnamefont {P.}~\bibnamefont {Pizzochero}}, \bibinfo {editor} {\bibfnamefont {D.~I.}\ \bibnamefont {Jones}}, \bibinfo {editor} {\bibfnamefont {N.}~\bibnamefont {Rea}},\ and\ \bibinfo {editor} {\bibfnamefont {I.}~\bibnamefont {Vida{\~n}a}},\ eds.,\ \href {https://doi.org/10.1007/978-3-319-97616-7} {\emph {\bibinfo {title} {{The Physics and Astrophysics of Neutron Stars}}}},\ \bibinfo {series} {Astrophysics and Space Science Library}, Vol.\ \bibinfo {volume} {457}\ (\bibinfo  {publisher} {Springer},\ \bibinfo {year} {2018})\BibitemShut {NoStop}%
\bibitem [{\citenamefont {Baym}\ \emph {et~al.}(2018)\citenamefont {Baym}, \citenamefont {Hatsuda}, \citenamefont {Kojo}, \citenamefont {Powell}, \citenamefont {Song},\ and\ \citenamefont {Takatsuka}}]{baym_hadrons_2018}%
  \BibitemOpen
  \bibfield  {author} {\bibinfo {author} {\bibfnamefont {G.}~\bibnamefont {Baym}}, \bibinfo {author} {\bibfnamefont {T.}~\bibnamefont {Hatsuda}}, \bibinfo {author} {\bibfnamefont {T.}~\bibnamefont {Kojo}}, \bibinfo {author} {\bibfnamefont {P.~D.}\ \bibnamefont {Powell}}, \bibinfo {author} {\bibfnamefont {Y.}~\bibnamefont {Song}},\ and\ \bibinfo {author} {\bibfnamefont {T.}~\bibnamefont {Takatsuka}},\ }\bibfield  {title} {\bibinfo {title} {From hadrons to quarks in neutron stars: A review},\ }\href {https://doi.org/10.1088/1361-6633/aaae14} {\bibfield  {journal} {\bibinfo  {journal} {Rep. Prog. Phys.}\ }\textbf {\bibinfo {volume} {81}},\ \bibinfo {pages} {056902} (\bibinfo {year} {2018})}\BibitemShut {NoStop}%
\bibitem [{\citenamefont {Burgio}\ \emph {et~al.}(2021)\citenamefont {Burgio}, \citenamefont {Schulze}, \citenamefont {Vida{\~n}a},\ and\ \citenamefont {Wei}}]{burgio_neutron_2021}%
  \BibitemOpen
  \bibfield  {author} {\bibinfo {author} {\bibfnamefont {G.~F.}\ \bibnamefont {Burgio}}, \bibinfo {author} {\bibfnamefont {H.~J.}\ \bibnamefont {Schulze}}, \bibinfo {author} {\bibfnamefont {I.}~\bibnamefont {Vida{\~n}a}},\ and\ \bibinfo {author} {\bibfnamefont {J.~B.}\ \bibnamefont {Wei}},\ }\bibfield  {title} {\bibinfo {title} {Neutron stars and the nuclear equation of state},\ }\href {https://doi.org/10.1016/j.ppnp.2021.103879} {\bibfield  {journal} {\bibinfo  {journal} {Progress in Particle and Nuclear Physics}\ }\textbf {\bibinfo {volume} {120}},\ \bibinfo {pages} {103879} (\bibinfo {year} {2021})}\BibitemShut {NoStop}%
\bibitem [{\citenamefont {Paschalidis}\ \emph {et~al.}(2018)\citenamefont {Paschalidis}, \citenamefont {Yagi}, \citenamefont {{Alvarez-Castillo}}, \citenamefont {Blaschke},\ and\ \citenamefont {Sedrakian}}]{paschalidis_implications_2018}%
  \BibitemOpen
  \bibfield  {author} {\bibinfo {author} {\bibfnamefont {V.}~\bibnamefont {Paschalidis}}, \bibinfo {author} {\bibfnamefont {K.}~\bibnamefont {Yagi}}, \bibinfo {author} {\bibfnamefont {D.}~\bibnamefont {{Alvarez-Castillo}}}, \bibinfo {author} {\bibfnamefont {D.~B.}\ \bibnamefont {Blaschke}},\ and\ \bibinfo {author} {\bibfnamefont {A.}~\bibnamefont {Sedrakian}},\ }\bibfield  {title} {\bibinfo {title} {Implications from {{GW170817}} and {{I-Love-Q}} relations for relativistic hybrid stars},\ }\href {https://doi.org/10.1103/PhysRevD.97.084038} {\bibfield  {journal} {\bibinfo  {journal} {Phys. Rev. D}\ }\textbf {\bibinfo {volume} {97}},\ \bibinfo {pages} {084038} (\bibinfo {year} {2018})}\BibitemShut {NoStop}%
\bibitem [{\citenamefont {Most}\ \emph {et~al.}(2019)\citenamefont {Most}, \citenamefont {Papenfort}, \citenamefont {Dexheimer}, \citenamefont {Hanauske}, \citenamefont {Schramm}, \citenamefont {St\"ocker},\ and\ \citenamefont {Rezzolla}}]{Most:2018eaw}%
  \BibitemOpen
  \bibfield  {author} {\bibinfo {author} {\bibfnamefont {E.~R.}\ \bibnamefont {Most}}, \bibinfo {author} {\bibfnamefont {L.~J.}\ \bibnamefont {Papenfort}}, \bibinfo {author} {\bibfnamefont {V.}~\bibnamefont {Dexheimer}}, \bibinfo {author} {\bibfnamefont {M.}~\bibnamefont {Hanauske}}, \bibinfo {author} {\bibfnamefont {S.}~\bibnamefont {Schramm}}, \bibinfo {author} {\bibfnamefont {H.}~\bibnamefont {St\"ocker}},\ and\ \bibinfo {author} {\bibfnamefont {L.}~\bibnamefont {Rezzolla}},\ }\bibfield  {title} {\bibinfo {title} {{Signatures of quark-hadron phase transitions in general-relativistic neutron-star mergers}},\ }\href {https://doi.org/10.1103/PhysRevLett.122.061101} {\bibfield  {journal} {\bibinfo  {journal} {Phys. Rev. Lett.}\ }\textbf {\bibinfo {volume} {122}},\ \bibinfo {pages} {061101} (\bibinfo {year} {2019})},\ \Eprint {https://arxiv.org/abs/1807.03684} {arXiv:1807.03684 [astro-ph.HE]} \BibitemShut {NoStop}%
\bibitem [{\citenamefont {Bauswein}\ \emph {et~al.}(2019)\citenamefont {Bauswein}, \citenamefont {Bastian}, \citenamefont {Blaschke}, \citenamefont {Chatziioannou}, \citenamefont {Clark}, \citenamefont {Fischer},\ and\ \citenamefont {Oertel}}]{bauswein_identifying_2019}%
  \BibitemOpen
  \bibfield  {author} {\bibinfo {author} {\bibfnamefont {A.}~\bibnamefont {Bauswein}}, \bibinfo {author} {\bibfnamefont {N.-U.~F.}\ \bibnamefont {Bastian}}, \bibinfo {author} {\bibfnamefont {D.~B.}\ \bibnamefont {Blaschke}}, \bibinfo {author} {\bibfnamefont {K.}~\bibnamefont {Chatziioannou}}, \bibinfo {author} {\bibfnamefont {J.~A.}\ \bibnamefont {Clark}}, \bibinfo {author} {\bibfnamefont {T.}~\bibnamefont {Fischer}},\ and\ \bibinfo {author} {\bibfnamefont {M.}~\bibnamefont {Oertel}},\ }\bibfield  {title} {\bibinfo {title} {Identifying a {{First-Order Phase Transition}} in {{Neutron-Star Mergers}} through {{Gravitational Waves}}},\ }\href {https://doi.org/10.1103/PhysRevLett.122.061102} {\bibfield  {journal} {\bibinfo  {journal} {Phys. Rev. Lett.}\ }\textbf {\bibinfo {volume} {122}},\ \bibinfo {pages} {061102} (\bibinfo {year} {2019})}\BibitemShut {NoStop}%
\bibitem [{\citenamefont {Weih}\ \emph {et~al.}(2020)\citenamefont {Weih}, \citenamefont {Hanauske},\ and\ \citenamefont {Rezzolla}}]{Weih:2019xvw}%
  \BibitemOpen
  \bibfield  {author} {\bibinfo {author} {\bibfnamefont {L.~R.}\ \bibnamefont {Weih}}, \bibinfo {author} {\bibfnamefont {M.}~\bibnamefont {Hanauske}},\ and\ \bibinfo {author} {\bibfnamefont {L.}~\bibnamefont {Rezzolla}},\ }\bibfield  {title} {\bibinfo {title} {{Postmerger Gravitational-Wave Signatures of Phase Transitions in Binary Mergers}},\ }\href {https://doi.org/10.1103/PhysRevLett.124.171103} {\bibfield  {journal} {\bibinfo  {journal} {Phys. Rev. Lett.}\ }\textbf {\bibinfo {volume} {124}},\ \bibinfo {pages} {171103} (\bibinfo {year} {2020})},\ \Eprint {https://arxiv.org/abs/1912.09340} {arXiv:1912.09340 [gr-qc]} \BibitemShut {NoStop}%
\bibitem [{\citenamefont {Prakash}\ \emph {et~al.}(2021)\citenamefont {Prakash}, \citenamefont {Radice}, \citenamefont {Logoteta}, \citenamefont {Perego}, \citenamefont {Nedora}, \citenamefont {Bombaci}, \citenamefont {Kashyap}, \citenamefont {Bernuzzi},\ and\ \citenamefont {Endrizzi}}]{prakash_signatures_2021}%
  \BibitemOpen
  \bibfield  {author} {\bibinfo {author} {\bibfnamefont {A.}~\bibnamefont {Prakash}}, \bibinfo {author} {\bibfnamefont {D.}~\bibnamefont {Radice}}, \bibinfo {author} {\bibfnamefont {D.}~\bibnamefont {Logoteta}}, \bibinfo {author} {\bibfnamefont {A.}~\bibnamefont {Perego}}, \bibinfo {author} {\bibfnamefont {V.}~\bibnamefont {Nedora}}, \bibinfo {author} {\bibfnamefont {I.}~\bibnamefont {Bombaci}}, \bibinfo {author} {\bibfnamefont {R.}~\bibnamefont {Kashyap}}, \bibinfo {author} {\bibfnamefont {S.}~\bibnamefont {Bernuzzi}},\ and\ \bibinfo {author} {\bibfnamefont {A.}~\bibnamefont {Endrizzi}},\ }\bibfield  {title} {\bibinfo {title} {Signatures of deconfined quark phases in binary neutron star mergers},\ }\href {https://doi.org/10.1103/PhysRevD.104.083029} {\bibfield  {journal} {\bibinfo  {journal} {Phys. Rev. D}\ }\textbf {\bibinfo {volume} {104}},\ \bibinfo {pages} {083029} (\bibinfo {year} {2021})}\BibitemShut {NoStop}%
\bibitem [{\citenamefont {Huang}\ \emph {et~al.}(2022)\citenamefont {Huang}, \citenamefont {Baiotti}, \citenamefont {Kojo}, \citenamefont {Takami}, \citenamefont {Sotani}, \citenamefont {Togashi}, \citenamefont {Hatsuda}, \citenamefont {Nagataki},\ and\ \citenamefont {Fan}}]{huang_merger_2022}%
  \BibitemOpen
  \bibfield  {author} {\bibinfo {author} {\bibfnamefont {Y.-J.}\ \bibnamefont {Huang}}, \bibinfo {author} {\bibfnamefont {L.}~\bibnamefont {Baiotti}}, \bibinfo {author} {\bibfnamefont {T.}~\bibnamefont {Kojo}}, \bibinfo {author} {\bibfnamefont {K.}~\bibnamefont {Takami}}, \bibinfo {author} {\bibfnamefont {H.}~\bibnamefont {Sotani}}, \bibinfo {author} {\bibfnamefont {H.}~\bibnamefont {Togashi}}, \bibinfo {author} {\bibfnamefont {T.}~\bibnamefont {Hatsuda}}, \bibinfo {author} {\bibfnamefont {S.}~\bibnamefont {Nagataki}},\ and\ \bibinfo {author} {\bibfnamefont {Y.-Z.}\ \bibnamefont {Fan}},\ }\bibfield  {title} {\bibinfo {title} {Merger and {{Postmerger}} of {{Binary Neutron Stars}} with a {{Quark-Hadron Crossover Equation}} of {{State}}},\ }\href {https://doi.org/10.1103/PhysRevLett.129.181101} {\bibfield  {journal} {\bibinfo  {journal} {Phys. Rev. Lett.}\ }\textbf {\bibinfo {volume} {129}},\ \bibinfo {pages} {181101} (\bibinfo {year} {2022})}\BibitemShut {NoStop}%
\bibitem [{\citenamefont {Fujimoto}\ \emph {et~al.}(2023)\citenamefont {Fujimoto}, \citenamefont {Fukushima}, \citenamefont {Hotokezaka},\ and\ \citenamefont {Kyutoku}}]{fujimoto_gravitational_2023}%
  \BibitemOpen
  \bibfield  {author} {\bibinfo {author} {\bibfnamefont {Y.}~\bibnamefont {Fujimoto}}, \bibinfo {author} {\bibfnamefont {K.}~\bibnamefont {Fukushima}}, \bibinfo {author} {\bibfnamefont {K.}~\bibnamefont {Hotokezaka}},\ and\ \bibinfo {author} {\bibfnamefont {K.}~\bibnamefont {Kyutoku}},\ }\bibfield  {title} {\bibinfo {title} {Gravitational {{Wave Signal}} for {{Quark Matter}} with {{Realistic Phase Transition}}},\ }\href {https://doi.org/10.1103/PhysRevLett.130.091404} {\bibfield  {journal} {\bibinfo  {journal} {Phys. Rev. Lett.}\ }\textbf {\bibinfo {volume} {130}},\ \bibinfo {pages} {091404} (\bibinfo {year} {2023})}\BibitemShut {NoStop}%
\bibitem [{\citenamefont {Ujevic}\ \emph {et~al.}(2023)\citenamefont {Ujevic}, \citenamefont {Gieg}, \citenamefont {Schianchi}, \citenamefont {Chaurasia}, \citenamefont {Tews},\ and\ \citenamefont {Dietrich}}]{ujevic_reverse_2023}%
  \BibitemOpen
  \bibfield  {author} {\bibinfo {author} {\bibfnamefont {M.}~\bibnamefont {Ujevic}}, \bibinfo {author} {\bibfnamefont {H.}~\bibnamefont {Gieg}}, \bibinfo {author} {\bibfnamefont {F.}~\bibnamefont {Schianchi}}, \bibinfo {author} {\bibfnamefont {S.~V.}\ \bibnamefont {Chaurasia}}, \bibinfo {author} {\bibfnamefont {I.}~\bibnamefont {Tews}},\ and\ \bibinfo {author} {\bibfnamefont {T.}~\bibnamefont {Dietrich}},\ }\bibfield  {title} {\bibinfo {title} {Reverse phase transitions in binary neutron-star systems with exotic-matter cores},\ }\href {https://doi.org/10.1103/PhysRevD.107.024025} {\bibfield  {journal} {\bibinfo  {journal} {Phys. Rev. D}\ }\textbf {\bibinfo {volume} {107}},\ \bibinfo {pages} {024025} (\bibinfo {year} {2023})}\BibitemShut {NoStop}%
\bibitem [{\citenamefont {Haque}\ \emph {et~al.}(2024)\citenamefont {Haque}, \citenamefont {Mallick},\ and\ \citenamefont {Thakur}}]{haque_effects_2024}%
  \BibitemOpen
  \bibfield  {author} {\bibinfo {author} {\bibfnamefont {S.}~\bibnamefont {Haque}}, \bibinfo {author} {\bibfnamefont {R.}~\bibnamefont {Mallick}},\ and\ \bibinfo {author} {\bibfnamefont {S.~K.}\ \bibnamefont {Thakur}},\ }\bibfield  {title} {\bibinfo {title} {Effects of onset of phase transition on binary neutron star mergers},\ }\href {https://doi.org/10.1093/mnras/stad3839} {\bibfield  {journal} {\bibinfo  {journal} {Mon Not R Astron Soc}\ }\textbf {\bibinfo {volume} {527}},\ \bibinfo {pages} {11575} (\bibinfo {year} {2024})}\BibitemShut {NoStop}%
\bibitem [{\citenamefont {Sagert}\ \emph {et~al.}(2009)\citenamefont {Sagert}, \citenamefont {Fischer}, \citenamefont {Hempel}, \citenamefont {Pagliara}, \citenamefont {{Schaffner-Bielich}}, \citenamefont {Mezzacappa}, \citenamefont {Thielemann},\ and\ \citenamefont {Liebend{\"o}rfer}}]{sagert_signals_2009}%
  \BibitemOpen
  \bibfield  {author} {\bibinfo {author} {\bibfnamefont {I.}~\bibnamefont {Sagert}}, \bibinfo {author} {\bibfnamefont {T.}~\bibnamefont {Fischer}}, \bibinfo {author} {\bibfnamefont {M.}~\bibnamefont {Hempel}}, \bibinfo {author} {\bibfnamefont {G.}~\bibnamefont {Pagliara}}, \bibinfo {author} {\bibfnamefont {J.}~\bibnamefont {{Schaffner-Bielich}}}, \bibinfo {author} {\bibfnamefont {A.}~\bibnamefont {Mezzacappa}}, \bibinfo {author} {\bibfnamefont {F.-K.}\ \bibnamefont {Thielemann}},\ and\ \bibinfo {author} {\bibfnamefont {M.}~\bibnamefont {Liebend{\"o}rfer}},\ }\bibfield  {title} {\bibinfo {title} {Signals of the {{QCD Phase Transition}} in {{Core-Collapse Supernovae}}},\ }\href {https://doi.org/10.1103/PhysRevLett.102.081101} {\bibfield  {journal} {\bibinfo  {journal} {Phys. Rev. Lett.}\ }\textbf {\bibinfo {volume} {102}},\ \bibinfo {pages} {081101} (\bibinfo {year} {2009})}\BibitemShut {NoStop}%
\bibitem [{\citenamefont {Zha}\ \emph {et~al.}(2021)\citenamefont {Zha}, \citenamefont {O'Connor},\ and\ \citenamefont {Da~Silva~Schneider}}]{zha_progenitor_2021}%
  \BibitemOpen
  \bibfield  {author} {\bibinfo {author} {\bibfnamefont {S.}~\bibnamefont {Zha}}, \bibinfo {author} {\bibfnamefont {E.~P.}\ \bibnamefont {O'Connor}},\ and\ \bibinfo {author} {\bibfnamefont {A.}~\bibnamefont {Da~Silva~Schneider}},\ }\bibfield  {title} {\bibinfo {title} {Progenitor {{Dependence}} of {{Hadron-quark Phase Transition}} in {{Failing Core-collapse Supernovae}}},\ }\href {https://doi.org/10.3847/1538-4357/abec4c} {\bibfield  {journal} {\bibinfo  {journal} {ApJ}\ }\textbf {\bibinfo {volume} {911}},\ \bibinfo {pages} {74} (\bibinfo {year} {2021})}\BibitemShut {NoStop}%
\bibitem [{\citenamefont {Kuroda}\ \emph {et~al.}(2022)\citenamefont {Kuroda}, \citenamefont {Fischer}, \citenamefont {Takiwaki},\ and\ \citenamefont {Kotake}}]{kuroda_core-collapse_2022}%
  \BibitemOpen
  \bibfield  {author} {\bibinfo {author} {\bibfnamefont {T.}~\bibnamefont {Kuroda}}, \bibinfo {author} {\bibfnamefont {T.}~\bibnamefont {Fischer}}, \bibinfo {author} {\bibfnamefont {T.}~\bibnamefont {Takiwaki}},\ and\ \bibinfo {author} {\bibfnamefont {K.}~\bibnamefont {Kotake}},\ }\bibfield  {title} {\bibinfo {title} {Core-collapse {{Supernova Simulations}} and the {{Formation}} of {{Neutron Stars}}, {{Hybrid Stars}}, and {{Black Holes}}},\ }\href {https://doi.org/10.3847/1538-4357/ac31a8} {\bibfield  {journal} {\bibinfo  {journal} {ApJ}\ }\textbf {\bibinfo {volume} {924}},\ \bibinfo {pages} {38} (\bibinfo {year} {2022})}\BibitemShut {NoStop}%
\bibitem [{\citenamefont {Jakobus}\ \emph {et~al.}(2022)\citenamefont {Jakobus}, \citenamefont {M{\"u}ller}, \citenamefont {Heger}, \citenamefont {Motornenko}, \citenamefont {Steinheimer},\ and\ \citenamefont {Stoecker}}]{jakobus_role_2022}%
  \BibitemOpen
  \bibfield  {author} {\bibinfo {author} {\bibfnamefont {P.}~\bibnamefont {Jakobus}}, \bibinfo {author} {\bibfnamefont {B.}~\bibnamefont {M{\"u}ller}}, \bibinfo {author} {\bibfnamefont {A.}~\bibnamefont {Heger}}, \bibinfo {author} {\bibfnamefont {A.}~\bibnamefont {Motornenko}}, \bibinfo {author} {\bibfnamefont {J.}~\bibnamefont {Steinheimer}},\ and\ \bibinfo {author} {\bibfnamefont {H.}~\bibnamefont {Stoecker}},\ }\bibfield  {title} {\bibinfo {title} {The role of the hadron-quark phase transition in core-collapse supernovae},\ }\href {https://doi.org/10.1093/mnras/stac2352} {\bibfield  {journal} {\bibinfo  {journal} {Mon Not R Astron Soc}\ }\textbf {\bibinfo {volume} {516}},\ \bibinfo {pages} {2554} (\bibinfo {year} {2022})}\BibitemShut {NoStop}%
\bibitem [{\citenamefont {Lonardoni}\ \emph {et~al.}(2020)\citenamefont {Lonardoni}, \citenamefont {Tews}, \citenamefont {Gandolfi},\ and\ \citenamefont {Carlson}}]{lonardoni_nuclear_2020}%
  \BibitemOpen
  \bibfield  {author} {\bibinfo {author} {\bibfnamefont {D.}~\bibnamefont {Lonardoni}}, \bibinfo {author} {\bibfnamefont {I.}~\bibnamefont {Tews}}, \bibinfo {author} {\bibfnamefont {S.}~\bibnamefont {Gandolfi}},\ and\ \bibinfo {author} {\bibfnamefont {J.}~\bibnamefont {Carlson}},\ }\bibfield  {title} {\bibinfo {title} {Nuclear and neutron-star matter from local chiral interactions},\ }\href {https://doi.org/10.1103/PhysRevResearch.2.022033} {\bibfield  {journal} {\bibinfo  {journal} {Phys. Rev. Research}\ }\textbf {\bibinfo {volume} {2}},\ \bibinfo {pages} {022033} (\bibinfo {year} {2020})}\BibitemShut {NoStop}%
\bibitem [{\citenamefont {Drischler}\ \emph {et~al.}(2021)\citenamefont {Drischler}, \citenamefont {Holt},\ and\ \citenamefont {Wellenhofer}}]{drischler_chiral_2021}%
  \BibitemOpen
  \bibfield  {author} {\bibinfo {author} {\bibfnamefont {C.}~\bibnamefont {Drischler}}, \bibinfo {author} {\bibfnamefont {J.~W.}\ \bibnamefont {Holt}},\ and\ \bibinfo {author} {\bibfnamefont {C.}~\bibnamefont {Wellenhofer}},\ }\bibfield  {title} {\bibinfo {title} {Chiral {{Effective Field Theory}} and the {{High-Density Nuclear Equation}} of {{State}}},\ }\href {https://doi.org/10.1146/annurev-nucl-102419-041903} {\bibfield  {journal} {\bibinfo  {journal} {Annual Review of Nuclear and Particle Science}\ }\textbf {\bibinfo {volume} {71}},\ \bibinfo {pages} {403} (\bibinfo {year} {2021})}\BibitemShut {NoStop}%
\bibitem [{\citenamefont {Adams}\ \emph {et~al.}(2005)\citenamefont {Adams} \emph {et~al.}}]{adams_experimental_2005}%
  \BibitemOpen
  \bibfield  {author} {\bibinfo {author} {\bibfnamefont {J.}~\bibnamefont {Adams}} \emph {et~al.},\ }\bibfield  {title} {\bibinfo {title} {Experimental and theoretical challenges in the search for the quark--gluon plasma: {{The STAR Collaboration}}'s critical assessment of the evidence from {{RHIC}} collisions},\ }\href {https://doi.org/10.1016/j.nuclphysa.2005.03.085} {\bibfield  {journal} {\bibinfo  {journal} {Nuclear Physics A}\ }\textbf {\bibinfo {volume} {757}},\ \bibinfo {pages} {102} (\bibinfo {year} {2005})}\BibitemShut {NoStop}%
\bibitem [{\citenamefont {Arsene}\ \emph {et~al.}(2005)\citenamefont {Arsene} \emph {et~al.}}]{arsene_quarkgluon_2005}%
  \BibitemOpen
  \bibfield  {author} {\bibinfo {author} {\bibfnamefont {I.}~\bibnamefont {Arsene}} \emph {et~al.},\ }\bibfield  {title} {\bibinfo {title} {Quark--gluon plasma and color glass condensate at {{RHIC}}? {{The}} perspective from the {{BRAHMS}} experiment},\ }\href {https://doi.org/10.1016/j.nuclphysa.2005.02.130} {\bibfield  {journal} {\bibinfo  {journal} {Nuclear Physics A}\ }\bibinfo {series} {First {{Three Years}} of {{Operation}} of {{RHIC}}},\ \textbf {\bibinfo {volume} {757}},\ \bibinfo {pages} {1} (\bibinfo {year} {2005})}\BibitemShut {NoStop}%
\bibitem [{\citenamefont {Back}\ \emph {et~al.}(2005)\citenamefont {Back} \emph {et~al.}}]{back_phobos_2005}%
  \BibitemOpen
  \bibfield  {author} {\bibinfo {author} {\bibfnamefont {B.~B.}\ \bibnamefont {Back}} \emph {et~al.},\ }\bibfield  {title} {\bibinfo {title} {The {{PHOBOS}} perspective on discoveries at {{RHIC}}},\ }\href {https://doi.org/10.1016/j.nuclphysa.2005.03.084} {\bibfield  {journal} {\bibinfo  {journal} {Nuclear Physics A}\ }\bibinfo {series} {First {{Three Years}} of {{Operation}} of {{RHIC}}},\ \textbf {\bibinfo {volume} {757}},\ \bibinfo {pages} {28} (\bibinfo {year} {2005})}\BibitemShut {NoStop}%
\bibitem [{\citenamefont {Adcox}\ \emph {et~al.}(2005)\citenamefont {Adcox} \emph {et~al.}}]{adcox_formation_2005}%
  \BibitemOpen
  \bibfield  {author} {\bibinfo {author} {\bibfnamefont {K.}~\bibnamefont {Adcox}} \emph {et~al.},\ }\bibfield  {title} {\bibinfo {title} {Formation of dense partonic matter in relativistic nucleus--nucleus collisions at {{RHIC}}: {{Experimental}} evaluation by the {{PHENIX Collaboration}}},\ }\href {https://doi.org/10.1016/j.nuclphysa.2005.03.086} {\bibfield  {journal} {\bibinfo  {journal} {Nuclear Physics A}\ }\bibinfo {series} {First {{Three Years}} of {{Operation}} of {{RHIC}}},\ \textbf {\bibinfo {volume} {757}},\ \bibinfo {pages} {184} (\bibinfo {year} {2005})}\BibitemShut {NoStop}%
\bibitem [{\citenamefont {Bors{\'a}nyi}\ \emph {et~al.}(2014)\citenamefont {Bors{\'a}nyi}, \citenamefont {Fodor}, \citenamefont {Hoelbling}, \citenamefont {Katz}, \citenamefont {Krieg},\ and\ \citenamefont {Szab{\'o}}}]{borsanyi_full_2014}%
  \BibitemOpen
  \bibfield  {author} {\bibinfo {author} {\bibfnamefont {S.}~\bibnamefont {Bors{\'a}nyi}}, \bibinfo {author} {\bibfnamefont {Z.}~\bibnamefont {Fodor}}, \bibinfo {author} {\bibfnamefont {C.}~\bibnamefont {Hoelbling}}, \bibinfo {author} {\bibfnamefont {S.~D.}\ \bibnamefont {Katz}}, \bibinfo {author} {\bibfnamefont {S.}~\bibnamefont {Krieg}},\ and\ \bibinfo {author} {\bibfnamefont {K.~K.}\ \bibnamefont {Szab{\'o}}},\ }\bibfield  {title} {\bibinfo {title} {Full result for the {{QCD}} equation of state with 2 + 1 flavors},\ }\href {https://doi.org/10.1016/j.physletb.2014.01.007} {\bibfield  {journal} {\bibinfo  {journal} {Physics Letters B}\ }\textbf {\bibinfo {volume} {730}},\ \bibinfo {pages} {99} (\bibinfo {year} {2014})}\BibitemShut {NoStop}%
\bibitem [{\citenamefont {Nagata}(2022)}]{nagata_finite-density_2022}%
  \BibitemOpen
  \bibfield  {author} {\bibinfo {author} {\bibfnamefont {K.}~\bibnamefont {Nagata}},\ }\bibfield  {title} {\bibinfo {title} {Finite-density lattice {{QCD}} and sign problem: {{Current}} status and open problems},\ }\href {https://doi.org/10.1016/j.ppnp.2022.103991} {\bibfield  {journal} {\bibinfo  {journal} {Progress in Particle and Nuclear Physics}\ }\textbf {\bibinfo {volume} {127}},\ \bibinfo {pages} {103991} (\bibinfo {year} {2022})}\BibitemShut {NoStop}%
\bibitem [{\citenamefont {{Most}}\ \emph {et~al.}(2023)\citenamefont {{Most}}, \citenamefont {{Motornenko}}, \citenamefont {{Steinheimer}}, \citenamefont {{Dexheimer}}, \citenamefont {{Hanauske}}, \citenamefont {{Rezzolla}},\ and\ \citenamefont {{Stoecker}}}]{Most2022e}%
  \BibitemOpen
  \bibfield  {author} {\bibinfo {author} {\bibfnamefont {E.~R.}\ \bibnamefont {{Most}}}, \bibinfo {author} {\bibfnamefont {A.}~\bibnamefont {{Motornenko}}}, \bibinfo {author} {\bibfnamefont {J.}~\bibnamefont {{Steinheimer}}}, \bibinfo {author} {\bibfnamefont {V.}~\bibnamefont {{Dexheimer}}}, \bibinfo {author} {\bibfnamefont {M.}~\bibnamefont {{Hanauske}}}, \bibinfo {author} {\bibfnamefont {L.}~\bibnamefont {{Rezzolla}}},\ and\ \bibinfo {author} {\bibfnamefont {H.}~\bibnamefont {{Stoecker}}},\ }\bibfield  {title} {\bibinfo {title} {{Probing neutron-star matter in the lab: Similarities and differences between binary mergers and heavy-ion collisions}},\ }\href {https://doi.org/10.1103/PhysRevD.107.043034} {\bibfield  {journal} {\bibinfo  {journal} {Phys. Rev. D}\ }\textbf {\bibinfo {volume} {107}},\ \bibinfo {eid} {043034} (\bibinfo {year} {2023})},\ \Eprint {https://arxiv.org/abs/2201.13150} {arXiv:2201.13150 [nucl-th]} \BibitemShut {NoStop}%
\bibitem [{\citenamefont {Mogliacci}\ \emph {et~al.}(2013)\citenamefont {Mogliacci}, \citenamefont {Andersen}, \citenamefont {Strickland}, \citenamefont {Su},\ and\ \citenamefont {Vuorinen}}]{mogliacci_equation_2013}%
  \BibitemOpen
  \bibfield  {author} {\bibinfo {author} {\bibfnamefont {S.}~\bibnamefont {Mogliacci}}, \bibinfo {author} {\bibfnamefont {J.~O.}\ \bibnamefont {Andersen}}, \bibinfo {author} {\bibfnamefont {M.}~\bibnamefont {Strickland}}, \bibinfo {author} {\bibfnamefont {N.}~\bibnamefont {Su}},\ and\ \bibinfo {author} {\bibfnamefont {A.}~\bibnamefont {Vuorinen}},\ }\bibfield  {title} {\bibinfo {title} {Equation of state of hot and dense {{QCD}}: Resummed perturbation theory confronts lattice data},\ }\href {https://doi.org/10.1007/JHEP12(2013)055} {\bibfield  {journal} {\bibinfo  {journal} {J. High Energ. Phys.}\ }\textbf {\bibinfo {volume} {2013}}\bibinfo  {number} { (12)},\ \bibinfo {pages} {55}}\BibitemShut {NoStop}%
\bibitem [{\citenamefont {Kurkela}\ and\ \citenamefont {Vuorinen}(2016)}]{kurkela_cool_2016}%
  \BibitemOpen
\bibfield  {number} {  }\bibfield  {author} {\bibinfo {author} {\bibfnamefont {A.}~\bibnamefont {Kurkela}}\ and\ \bibinfo {author} {\bibfnamefont {A.}~\bibnamefont {Vuorinen}},\ }\bibfield  {title} {\bibinfo {title} {Cool {{Quark Matter}}},\ }\href {https://doi.org/10.1103/PhysRevLett.117.042501} {\bibfield  {journal} {\bibinfo  {journal} {Phys. Rev. Lett.}\ }\textbf {\bibinfo {volume} {117}},\ \bibinfo {pages} {042501} (\bibinfo {year} {2016})}\BibitemShut {NoStop}%
\bibitem [{\citenamefont {Alford}\ and\ \citenamefont {Sedrakian}(2017)}]{alford_compact_2017}%
  \BibitemOpen
  \bibfield  {author} {\bibinfo {author} {\bibfnamefont {M.}~\bibnamefont {Alford}}\ and\ \bibinfo {author} {\bibfnamefont {A.}~\bibnamefont {Sedrakian}},\ }\bibfield  {title} {\bibinfo {title} {Compact {{Stars}} with {{Sequential QCD Phase Transitions}}},\ }\href {https://doi.org/10.1103/PhysRevLett.119.161104} {\bibfield  {journal} {\bibinfo  {journal} {Phys. Rev. Lett.}\ }\textbf {\bibinfo {volume} {119}},\ \bibinfo {pages} {161104} (\bibinfo {year} {2017})}\BibitemShut {NoStop}%
\bibitem [{\citenamefont {Haque}\ and\ \citenamefont {Strickland}(2021)}]{haque_next--next-leading-order_2021}%
  \BibitemOpen
  \bibfield  {author} {\bibinfo {author} {\bibfnamefont {N.}~\bibnamefont {Haque}}\ and\ \bibinfo {author} {\bibfnamefont {M.}~\bibnamefont {Strickland}},\ }\bibfield  {title} {\bibinfo {title} {Next-to-next-to leading-order hard-thermal-loop perturbation-theory predictions for the curvature of the {{QCD}} phase transition line},\ }\href {https://doi.org/10.1103/PhysRevC.103.L031901} {\bibfield  {journal} {\bibinfo  {journal} {Phys. Rev. C}\ }\textbf {\bibinfo {volume} {103}},\ \bibinfo {pages} {L031901} (\bibinfo {year} {2021})}\BibitemShut {NoStop}%
\bibitem [{\citenamefont {Christian}\ and\ \citenamefont {{Schaffner-Bielich}}(2022)}]{christian_confirming_2022}%
  \BibitemOpen
  \bibfield  {author} {\bibinfo {author} {\bibfnamefont {J.-E.}\ \bibnamefont {Christian}}\ and\ \bibinfo {author} {\bibfnamefont {J.}~\bibnamefont {{Schaffner-Bielich}}},\ }\bibfield  {title} {\bibinfo {title} {Confirming the {{Existence}} of {{Twin Stars}} in a {{NICER Way}}},\ }\href {https://doi.org/10.3847/1538-4357/ac75cf} {\bibfield  {journal} {\bibinfo  {journal} {ApJ}\ }\textbf {\bibinfo {volume} {935}},\ \bibinfo {pages} {122} (\bibinfo {year} {2022})}\BibitemShut {NoStop}%
\bibitem [{\citenamefont {Gorda}\ \emph {et~al.}(2023)\citenamefont {Gorda}, \citenamefont {Paatelainen}, \citenamefont {S{\"a}ppi},\ and\ \citenamefont {Sepp{\"a}nen}}]{gorda_equation_2023}%
  \BibitemOpen
  \bibfield  {author} {\bibinfo {author} {\bibfnamefont {T.}~\bibnamefont {Gorda}}, \bibinfo {author} {\bibfnamefont {R.}~\bibnamefont {Paatelainen}}, \bibinfo {author} {\bibfnamefont {S.}~\bibnamefont {S{\"a}ppi}},\ and\ \bibinfo {author} {\bibfnamefont {K.}~\bibnamefont {Sepp{\"a}nen}},\ }\bibfield  {title} {\bibinfo {title} {Equation of {{State}} of {{Cold Quark Matter}} to {{O}} ( {$\alpha$} s 3 ln {$\alpha$} s )},\ }\href {https://doi.org/10.1103/PhysRevLett.131.181902} {\bibfield  {journal} {\bibinfo  {journal} {Phys. Rev. Lett.}\ }\textbf {\bibinfo {volume} {131}},\ \bibinfo {pages} {181902} (\bibinfo {year} {2023})}\BibitemShut {NoStop}%
\bibitem [{\citenamefont {Oertel}\ \emph {et~al.}(2017)\citenamefont {Oertel}, \citenamefont {Hempel}, \citenamefont {Kl{\"a}hn},\ and\ \citenamefont {Typel}}]{oertel_equations_2017}%
  \BibitemOpen
  \bibfield  {author} {\bibinfo {author} {\bibfnamefont {M.}~\bibnamefont {Oertel}}, \bibinfo {author} {\bibfnamefont {M.}~\bibnamefont {Hempel}}, \bibinfo {author} {\bibfnamefont {T.}~\bibnamefont {Kl{\"a}hn}},\ and\ \bibinfo {author} {\bibfnamefont {S.}~\bibnamefont {Typel}},\ }\bibfield  {title} {\bibinfo {title} {Equations of state for supernovae and compact stars},\ }\href {https://doi.org/10.1103/RevModPhys.89.015007} {\bibfield  {journal} {\bibinfo  {journal} {Rev. Mod. Phys.}\ }\textbf {\bibinfo {volume} {89}},\ \bibinfo {pages} {015007} (\bibinfo {year} {2017})}\BibitemShut {NoStop}%
\bibitem [{\citenamefont {Blacker}\ \emph {et~al.}(2020)\citenamefont {Blacker}, \citenamefont {Bastian}, \citenamefont {Bauswein}, \citenamefont {Blaschke}, \citenamefont {Fischer}, \citenamefont {Oertel}, \citenamefont {Soultanis},\ and\ \citenamefont {Typel}}]{blacker_constraining_2020}%
  \BibitemOpen
  \bibfield  {author} {\bibinfo {author} {\bibfnamefont {S.}~\bibnamefont {Blacker}}, \bibinfo {author} {\bibfnamefont {N.-U.~F.}\ \bibnamefont {Bastian}}, \bibinfo {author} {\bibfnamefont {A.}~\bibnamefont {Bauswein}}, \bibinfo {author} {\bibfnamefont {D.~B.}\ \bibnamefont {Blaschke}}, \bibinfo {author} {\bibfnamefont {T.}~\bibnamefont {Fischer}}, \bibinfo {author} {\bibfnamefont {M.}~\bibnamefont {Oertel}}, \bibinfo {author} {\bibfnamefont {T.}~\bibnamefont {Soultanis}},\ and\ \bibinfo {author} {\bibfnamefont {S.}~\bibnamefont {Typel}},\ }\bibfield  {title} {\bibinfo {title} {Constraining the onset density of the hadron-quark phase transition with gravitational-wave observations},\ }\href {https://doi.org/10.1103/PhysRevD.102.123023} {\bibfield  {journal} {\bibinfo  {journal} {Phys. Rev. D}\ }\textbf {\bibinfo {volume} {102}},\ \bibinfo {pages} {123023} (\bibinfo {year} {2020})}\BibitemShut {NoStop}%
\bibitem [{\citenamefont {Dore}\ \emph {et~al.}(2020)\citenamefont {Dore}, \citenamefont {{Noronha-Hostler}},\ and\ \citenamefont {McLaughlin}}]{dore_far--equilibrium_2020}%
  \BibitemOpen
  \bibfield  {author} {\bibinfo {author} {\bibfnamefont {T.}~\bibnamefont {Dore}}, \bibinfo {author} {\bibfnamefont {J.}~\bibnamefont {{Noronha-Hostler}}},\ and\ \bibinfo {author} {\bibfnamefont {E.}~\bibnamefont {McLaughlin}},\ }\bibfield  {title} {\bibinfo {title} {Far-from-equilibrium search for the {{QCD}} critical point},\ }\href {https://doi.org/10.1103/PhysRevD.102.074017} {\bibfield  {journal} {\bibinfo  {journal} {Phys. Rev. D}\ }\textbf {\bibinfo {volume} {102}},\ \bibinfo {pages} {074017} (\bibinfo {year} {2020})}\BibitemShut {NoStop}%
\bibitem [{\citenamefont {Annala}\ \emph {et~al.}(2020)\citenamefont {Annala}, \citenamefont {Gorda}, \citenamefont {Kurkela}, \citenamefont {N{\"a}ttil{\"a}},\ and\ \citenamefont {Vuorinen}}]{annala_evidence_2020}%
  \BibitemOpen
  \bibfield  {author} {\bibinfo {author} {\bibfnamefont {E.}~\bibnamefont {Annala}}, \bibinfo {author} {\bibfnamefont {T.}~\bibnamefont {Gorda}}, \bibinfo {author} {\bibfnamefont {A.}~\bibnamefont {Kurkela}}, \bibinfo {author} {\bibfnamefont {J.}~\bibnamefont {N{\"a}ttil{\"a}}},\ and\ \bibinfo {author} {\bibfnamefont {A.}~\bibnamefont {Vuorinen}},\ }\bibfield  {title} {\bibinfo {title} {Evidence for quark-matter cores in massive neutron stars},\ }\href {https://doi.org/10.1038/s41567-020-0914-9} {\bibfield  {journal} {\bibinfo  {journal} {Nat. Phys.}\ }\textbf {\bibinfo {volume} {16}},\ \bibinfo {pages} {907} (\bibinfo {year} {2020})}\BibitemShut {NoStop}%
\bibitem [{\citenamefont {Verma}\ \emph {et~al.}(2025)\citenamefont {Verma}, \citenamefont {Saha}, \citenamefont {Malik},\ and\ \citenamefont {Mallick}}]{verma_probing_2025}%
  \BibitemOpen
  \bibfield  {author} {\bibinfo {author} {\bibfnamefont {A.}~\bibnamefont {Verma}}, \bibinfo {author} {\bibfnamefont {A.~K.}\ \bibnamefont {Saha}}, \bibinfo {author} {\bibfnamefont {T.}~\bibnamefont {Malik}},\ and\ \bibinfo {author} {\bibfnamefont {R.}~\bibnamefont {Mallick}},\ }\bibfield  {title} {\bibinfo {title} {Probing the {{Internal Structure}} of {{Neutron Stars}}: {{A Comparative Analysis}} of {{Three Different Classes}} of {{Equations}} of {{State}}},\ }\href {https://doi.org/10.3847/1538-4357/ade9a2} {\bibfield  {journal} {\bibinfo  {journal} {ApJ}\ }\textbf {\bibinfo {volume} {988}},\ \bibinfo {pages} {258} (\bibinfo {year} {2025})}\BibitemShut {NoStop}%
\bibitem [{\citenamefont {Glendenning}(2001)}]{glendenning_phase_2001}%
  \BibitemOpen
  \bibfield  {author} {\bibinfo {author} {\bibfnamefont {N.~K.}\ \bibnamefont {Glendenning}},\ }\bibfield  {title} {\bibinfo {title} {Phase transitions and crystalline structures in neutron star cores},\ }\href {https://doi.org/10.1016/S0370-1573(00)00080-6} {\bibfield  {journal} {\bibinfo  {journal} {Physics Reports}\ }\textbf {\bibinfo {volume} {342}},\ \bibinfo {pages} {393} (\bibinfo {year} {2001})}\BibitemShut {NoStop}%
\bibitem [{\citenamefont {Alford}\ \emph {et~al.}(2013)\citenamefont {Alford}, \citenamefont {Han},\ and\ \citenamefont {Prakash}}]{alford_generic_2013}%
  \BibitemOpen
  \bibfield  {author} {\bibinfo {author} {\bibfnamefont {M.~G.}\ \bibnamefont {Alford}}, \bibinfo {author} {\bibfnamefont {S.}~\bibnamefont {Han}},\ and\ \bibinfo {author} {\bibfnamefont {M.}~\bibnamefont {Prakash}},\ }\bibfield  {title} {\bibinfo {title} {Generic conditions for stable hybrid stars},\ }\href {https://doi.org/10.1103/PhysRevD.88.083013} {\bibfield  {journal} {\bibinfo  {journal} {Phys. Rev. D}\ }\textbf {\bibinfo {volume} {88}},\ \bibinfo {pages} {083013} (\bibinfo {year} {2013})}\BibitemShut {NoStop}%
\bibitem [{\citenamefont {Christian}\ \emph {et~al.}(2018)\citenamefont {Christian}, \citenamefont {Zacchi},\ and\ \citenamefont {{Schaffner-Bielich}}}]{christian_classifications_2018}%
  \BibitemOpen
  \bibfield  {author} {\bibinfo {author} {\bibfnamefont {J.-E.}\ \bibnamefont {Christian}}, \bibinfo {author} {\bibfnamefont {A.}~\bibnamefont {Zacchi}},\ and\ \bibinfo {author} {\bibfnamefont {J.}~\bibnamefont {{Schaffner-Bielich}}},\ }\bibfield  {title} {\bibinfo {title} {Classifications of twin star solutions for a constant speed of sound parameterized equation of state},\ }\href {https://doi.org/10.1140/epja/i2018-12472-y} {\bibfield  {journal} {\bibinfo  {journal} {Eur. Phys. J. A}\ }\textbf {\bibinfo {volume} {54}},\ \bibinfo {pages} {28} (\bibinfo {year} {2018})}\BibitemShut {NoStop}%
\bibitem [{\citenamefont {Monta{\~n}a}\ \emph {et~al.}(2019)\citenamefont {Monta{\~n}a}, \citenamefont {Tol{\'o}s}, \citenamefont {Hanauske},\ and\ \citenamefont {Rezzolla}}]{montana_constraining_2019}%
  \BibitemOpen
  \bibfield  {author} {\bibinfo {author} {\bibfnamefont {G.}~\bibnamefont {Monta{\~n}a}}, \bibinfo {author} {\bibfnamefont {L.}~\bibnamefont {Tol{\'o}s}}, \bibinfo {author} {\bibfnamefont {M.}~\bibnamefont {Hanauske}},\ and\ \bibinfo {author} {\bibfnamefont {L.}~\bibnamefont {Rezzolla}},\ }\bibfield  {title} {\bibinfo {title} {Constraining twin stars with {{GW170817}}},\ }\href {https://doi.org/10.1103/PhysRevD.99.103009} {\bibfield  {journal} {\bibinfo  {journal} {Phys. Rev. D}\ }\textbf {\bibinfo {volume} {99}},\ \bibinfo {pages} {103009} (\bibinfo {year} {2019})}\BibitemShut {NoStop}%
\bibitem [{\citenamefont {Haque}\ \emph {et~al.}(2026)\citenamefont {Haque}, \citenamefont {Shinde}, \citenamefont {Saha}, \citenamefont {Malik},\ and\ \citenamefont {Mallick}}]{haque_investigating_2026}%
  \BibitemOpen
  \bibfield  {author} {\bibinfo {author} {\bibfnamefont {S.}~\bibnamefont {Haque}}, \bibinfo {author} {\bibfnamefont {A.}~\bibnamefont {Shinde}}, \bibinfo {author} {\bibfnamefont {A.~K.}\ \bibnamefont {Saha}}, \bibinfo {author} {\bibfnamefont {T.}~\bibnamefont {Malik}},\ and\ \bibinfo {author} {\bibfnamefont {R.}~\bibnamefont {Mallick}},\ }\href {https://doi.org/10.48550/arXiv.2601.16674} {\bibinfo {title} {Investigating {{Twin Star Equation}} of {{States}} in {{Light}} of {{Recent Astrophysical Observations}}}} (\bibinfo {year} {2026}),\ \Eprint {https://arxiv.org/abs/2601.16674} {arXiv:2601.16674 [astro-ph]} \BibitemShut {NoStop}%
\bibitem [{\citenamefont {Font}\ \emph {et~al.}(2002)\citenamefont {Font}, \citenamefont {Goodale}, \citenamefont {Iyer}, \citenamefont {Miller}, \citenamefont {Rezzolla}, \citenamefont {Seidel}, \citenamefont {Stergioulas}, \citenamefont {Suen},\ and\ \citenamefont {Tobias}}]{font_three-dimensional_2002}%
  \BibitemOpen
  \bibfield  {author} {\bibinfo {author} {\bibfnamefont {J.~A.}\ \bibnamefont {Font}}, \bibinfo {author} {\bibfnamefont {T.}~\bibnamefont {Goodale}}, \bibinfo {author} {\bibfnamefont {S.}~\bibnamefont {Iyer}}, \bibinfo {author} {\bibfnamefont {M.}~\bibnamefont {Miller}}, \bibinfo {author} {\bibfnamefont {L.}~\bibnamefont {Rezzolla}}, \bibinfo {author} {\bibfnamefont {E.}~\bibnamefont {Seidel}}, \bibinfo {author} {\bibfnamefont {N.}~\bibnamefont {Stergioulas}}, \bibinfo {author} {\bibfnamefont {W.-M.}\ \bibnamefont {Suen}},\ and\ \bibinfo {author} {\bibfnamefont {M.}~\bibnamefont {Tobias}},\ }\bibfield  {title} {\bibinfo {title} {Three-dimensional numerical general relativistic hydrodynamics. {{II}}. {{Long-term}} dynamics of single relativistic stars},\ }\href {https://doi.org/10.1103/PhysRevD.65.084024} {\bibfield  {journal} {\bibinfo  {journal} {Phys. Rev. D}\ }\textbf {\bibinfo {volume} {65}},\ \bibinfo {pages} {084024} (\bibinfo {year} {2002})}\BibitemShut {NoStop}%
\bibitem [{\citenamefont {Dimmelmeier}\ \emph {et~al.}(2009)\citenamefont {Dimmelmeier}, \citenamefont {Bejger}, \citenamefont {Haensel},\ and\ \citenamefont {Zdunik}}]{dimmelmeier_dynamic_2009}%
  \BibitemOpen
  \bibfield  {author} {\bibinfo {author} {\bibfnamefont {H.}~\bibnamefont {Dimmelmeier}}, \bibinfo {author} {\bibfnamefont {M.}~\bibnamefont {Bejger}}, \bibinfo {author} {\bibfnamefont {P.}~\bibnamefont {Haensel}},\ and\ \bibinfo {author} {\bibfnamefont {J.~L.}\ \bibnamefont {Zdunik}},\ }\bibfield  {title} {\bibinfo {title} {Dynamic migration of rotating neutron stars due to a phase transition instability},\ }\href {https://doi.org/10.1111/j.1365-2966.2009.14891.x} {\bibfield  {journal} {\bibinfo  {journal} {Mon Not R Astron Soc}\ }\textbf {\bibinfo {volume} {396}},\ \bibinfo {pages} {2269} (\bibinfo {year} {2009})}\BibitemShut {NoStop}%
\bibitem [{\citenamefont {{Hanauske}}\ \emph {et~al.}(2018)\citenamefont {{Hanauske}}, \citenamefont {{Yilmaz, Zekiye Simay}}, \citenamefont {{Mitropoulos, Christina}}, \citenamefont {{Rezzolla, Luciano}},\ and\ \citenamefont {{St{\"o}cker, Horst}}}]{Hanauske2018}%
  \BibitemOpen
  \bibfield  {author} {\bibinfo {author} {\bibfnamefont {M.}~\bibnamefont {{Hanauske}}}, \bibinfo {author} {\bibnamefont {{Yilmaz, Zekiye Simay}}}, \bibinfo {author} {\bibnamefont {{Mitropoulos, Christina}}}, \bibinfo {author} {\bibnamefont {{Rezzolla, Luciano}}},\ and\ \bibinfo {author} {\bibnamefont {{St{\"o}cker, Horst}}},\ }\bibfield  {title} {\bibinfo {title} {Gravitational waves from binary compact star mergers in the context of strange matter},\ }\href {https://doi.org/10.1051/epjconf/201817120004} {\bibfield  {journal} {\bibinfo  {journal} {EPJ Web Conf.}\ }\textbf {\bibinfo {volume} {171}},\ \bibinfo {pages} {20004} (\bibinfo {year} {2018})}\BibitemShut {NoStop}%
\bibitem [{\citenamefont {Espino}\ and\ \citenamefont {Paschalidis}(2022)}]{espino_fate_2022}%
  \BibitemOpen
  \bibfield  {author} {\bibinfo {author} {\bibfnamefont {P.~L.}\ \bibnamefont {Espino}}\ and\ \bibinfo {author} {\bibfnamefont {V.}~\bibnamefont {Paschalidis}},\ }\bibfield  {title} {\bibinfo {title} {Fate of twin stars on the unstable branch: {{Implications}} for the formation of twin stars},\ }\href {https://doi.org/10.1103/PhysRevD.105.043014} {\bibfield  {journal} {\bibinfo  {journal} {Phys. Rev. D}\ }\textbf {\bibinfo {volume} {105}},\ \bibinfo {pages} {043014} (\bibinfo {year} {2022})}\BibitemShut {NoStop}%
\bibitem [{\citenamefont {Huang}\ \emph {et~al.}(2025)\citenamefont {Huang}, \citenamefont {Zha}, \citenamefont {Chu}, \citenamefont {O'Connor},\ and\ \citenamefont {Chen}}]{huang_phase-transition-induced_2025}%
  \BibitemOpen
  \bibfield  {author} {\bibinfo {author} {\bibfnamefont {X.-R.}\ \bibnamefont {Huang}}, \bibinfo {author} {\bibfnamefont {S.}~\bibnamefont {Zha}}, \bibinfo {author} {\bibfnamefont {M.-c.}\ \bibnamefont {Chu}}, \bibinfo {author} {\bibfnamefont {E.~P.}\ \bibnamefont {O'Connor}},\ and\ \bibinfo {author} {\bibfnamefont {L.-W.}\ \bibnamefont {Chen}},\ }\bibfield  {title} {\bibinfo {title} {Phase-transition-induced {{Collapse}} of {{Proto-compact Stars}} and {{Its Implication}} for {{Supernova Explosions}}},\ }\href {https://doi.org/10.3847/1538-4357/ada146} {\bibfield  {journal} {\bibinfo  {journal} {ApJ}\ }\textbf {\bibinfo {volume} {979}},\ \bibinfo {pages} {151} (\bibinfo {year} {2025})}\BibitemShut {NoStop}%
\bibitem [{\citenamefont {{Rezzolla}}\ and\ \citenamefont {{Zanotti}}(2013)}]{Rezzolla_book:2013}%
  \BibitemOpen
  \bibfield  {author} {\bibinfo {author} {\bibfnamefont {L.}~\bibnamefont {{Rezzolla}}}\ and\ \bibinfo {author} {\bibfnamefont {O.}~\bibnamefont {{Zanotti}}},\ }\href {https://doi.org/10.1093/acprof:oso/9780198528906.001.0001} {\emph {\bibinfo {title} {{Relativistic Hydrodynamics}}}}\ (\bibinfo  {publisher} {Oxford University Press},\ \bibinfo {year} {2013})\BibitemShut {NoStop}%
\bibitem [{\citenamefont {Naseri}\ \emph {et~al.}(2024)\citenamefont {Naseri}, \citenamefont {Bozzola},\ and\ \citenamefont {Paschalidis}}]{naseri_exploring_2024}%
  \BibitemOpen
  \bibfield  {author} {\bibinfo {author} {\bibfnamefont {M.}~\bibnamefont {Naseri}}, \bibinfo {author} {\bibfnamefont {G.}~\bibnamefont {Bozzola}},\ and\ \bibinfo {author} {\bibfnamefont {V.}~\bibnamefont {Paschalidis}},\ }\bibfield  {title} {\bibinfo {title} {Exploring pathways to forming twin stars},\ }\href {https://doi.org/10.1103/PhysRevD.110.044037} {\bibfield  {journal} {\bibinfo  {journal} {Phys. Rev. D}\ }\textbf {\bibinfo {volume} {110}},\ \bibinfo {pages} {044037} (\bibinfo {year} {2024})}\BibitemShut {NoStop}%
\bibitem [{\citenamefont {{Takami}}\ \emph {et~al.}(2011)\citenamefont {{Takami}}, \citenamefont {{Rezzolla}},\ and\ \citenamefont {{Yoshida}}}]{Takami:2011}%
  \BibitemOpen
  \bibfield  {author} {\bibinfo {author} {\bibfnamefont {K.}~\bibnamefont {{Takami}}}, \bibinfo {author} {\bibfnamefont {L.}~\bibnamefont {{Rezzolla}}},\ and\ \bibinfo {author} {\bibfnamefont {S.}~\bibnamefont {{Yoshida}}},\ }\bibfield  {title} {\bibinfo {title} {{A quasi-radial stability criterion for rotating relativistic stars}},\ }\href {https://doi.org/10.1111/j.1745-3933.2011.01085.x} {\bibfield  {journal} {\bibinfo  {journal} {Mon. Not. R. Astron. Soc.}\ }\textbf {\bibinfo {volume} {416}},\ \bibinfo {pages} {L1} (\bibinfo {year} {2011})},\ \Eprint {https://arxiv.org/abs/1105.3069} {arXiv:1105.3069 [gr-qc]} \BibitemShut {NoStop}%
\bibitem [{\citenamefont {{Garibay}}\ \emph {et~al.}(2026)\citenamefont {{Garibay}}, \citenamefont {{Ecker}},\ and\ \citenamefont {{Rezzolla}}}]{Garibay2026}%
  \BibitemOpen
  \bibfield  {author} {\bibinfo {author} {\bibfnamefont {I.}~\bibnamefont {{Garibay}}}, \bibinfo {author} {\bibfnamefont {C.}~\bibnamefont {{Ecker}}},\ and\ \bibinfo {author} {\bibfnamefont {L.}~\bibnamefont {{Rezzolla}}},\ }\bibfield  {title} {\bibinfo {title} {{General gravitational properties of neutron stars: curvature invariants, binding energy, and trace anomaly}},\ }\href {https://doi.org/10.48550/arXiv.2601.07931} {\bibfield  {journal} {\bibinfo  {journal} {arXiv e-prints}\ ,\ \bibinfo {eid} {arXiv:2601.07931}} (\bibinfo {year} {2026})},\ \Eprint {https://arxiv.org/abs/2601.07931} {arXiv:2601.07931 [gr-qc]} \BibitemShut {NoStop}%
\bibitem [{Note1()}]{Note1}%
  \BibitemOpen
  \bibinfo {note} {We recall that a quasi-universal relation can be used to relate $M$ and ${M}_\protect \mathrm {b}$ analytically~\cite {Timmes1996, Garibay2026}.}\BibitemShut {Stop}%
\bibitem [{\citenamefont {O'Connor}\ and\ \citenamefont {Ott}(2010)}]{oconnor_new_2010}%
  \BibitemOpen
  \bibfield  {author} {\bibinfo {author} {\bibfnamefont {E.}~\bibnamefont {O'Connor}}\ and\ \bibinfo {author} {\bibfnamefont {C.~D.}\ \bibnamefont {Ott}},\ }\bibfield  {title} {\bibinfo {title} {A new open-source code for spherically symmetric stellar collapse to neutron stars and black holes},\ }\href {https://doi.org/10.1088/0264-9381/27/11/114103} {\bibfield  {journal} {\bibinfo  {journal} {Class. Quantum Grav.}\ }\textbf {\bibinfo {volume} {27}},\ \bibinfo {pages} {114103} (\bibinfo {year} {2010})}\BibitemShut {NoStop}%
\bibitem [{\citenamefont {{Radice}}\ \emph {et~al.}(2014)\citenamefont {{Radice}}, \citenamefont {{Rezzolla}},\ and\ \citenamefont {{Galeazzi}}}]{Radice2013c}%
  \BibitemOpen
  \bibfield  {author} {\bibinfo {author} {\bibfnamefont {D.}~\bibnamefont {{Radice}}}, \bibinfo {author} {\bibfnamefont {L.}~\bibnamefont {{Rezzolla}}},\ and\ \bibinfo {author} {\bibfnamefont {F.}~\bibnamefont {{Galeazzi}}},\ }\bibfield  {title} {\bibinfo {title} {{High-order fully general-relativistic hydrodynamics: new approaches and tests}},\ }\href {https://doi.org/10.1088/0264-9381/31/7/075012} {\bibfield  {journal} {\bibinfo  {journal} {Class. Quantum Grav.}\ }\textbf {\bibinfo {volume} {31}},\ \bibinfo {eid} {075012} (\bibinfo {year} {2014})},\ \Eprint {https://arxiv.org/abs/1312.5004} {arXiv:1312.5004 [gr-qc]} \BibitemShut {NoStop}%
\bibitem [{\citenamefont {Shashank}\ \emph {et~al.}(2023)\citenamefont {Shashank}, \citenamefont {Nouri},\ and\ \citenamefont {Gupta}}]{shashank_f_2023}%
  \BibitemOpen
  \bibfield  {author} {\bibinfo {author} {\bibfnamefont {S.}~\bibnamefont {Shashank}}, \bibinfo {author} {\bibfnamefont {F.~H.}\ \bibnamefont {Nouri}},\ and\ \bibinfo {author} {\bibfnamefont {A.}~\bibnamefont {Gupta}},\ }\bibfield  {title} {\bibinfo {title} {F -mode oscillations of compact stars with realistic equations of state in dynamical spacetime},\ }\href {https://doi.org/10.1016/j.newast.2023.102067} {\bibfield  {journal} {\bibinfo  {journal} {New Astronomy}\ }\textbf {\bibinfo {volume} {104}},\ \bibinfo {pages} {102067} (\bibinfo {year} {2023})}\BibitemShut {NoStop}%
\bibitem [{\citenamefont {Pierre~Jacques}\ \emph {et~al.}(2025)\citenamefont {Pierre~Jacques}, \citenamefont {Cupp}, \citenamefont {Werneck}, \citenamefont {Tootle}, \citenamefont {Babiuc~Hamilton},\ and\ \citenamefont {Etienne}}]{pierre_jacques_general_2025}%
  \BibitemOpen
  \bibfield  {author} {\bibinfo {author} {\bibfnamefont {T.}~\bibnamefont {Pierre~Jacques}}, \bibinfo {author} {\bibfnamefont {S.}~\bibnamefont {Cupp}}, \bibinfo {author} {\bibfnamefont {L.~R.}\ \bibnamefont {Werneck}}, \bibinfo {author} {\bibfnamefont {S.~D.}\ \bibnamefont {Tootle}}, \bibinfo {author} {\bibfnamefont {M.~C.}\ \bibnamefont {Babiuc~Hamilton}},\ and\ \bibinfo {author} {\bibfnamefont {Z.~B.}\ \bibnamefont {Etienne}},\ }\bibfield  {title} {\bibinfo {title} {General relativistic hydrodynamics code for dynamical spacetimes with curvilinear coordinates, tabulated equations of state, and neutrino physics},\ }\href {https://doi.org/10.1103/hc9l-1thx} {\bibfield  {journal} {\bibinfo  {journal} {Phys. Rev. D}\ }\textbf {\bibinfo {volume} {112}},\ \bibinfo {pages} {084044} (\bibinfo {year} {2025})}\BibitemShut {NoStop}%
\bibitem [{Note2()}]{Note2}%
  \BibitemOpen
  \bibinfo {note} {We note that the equilibrium is towards slightly larger central rest-mass density values because the numerical import naturally introduces a difference in the equilibrium model~\cite [see][for a first discussion of this issue]{baiotti_three-dimensional_2005}.}\BibitemShut {Stop}%
\bibitem [{\citenamefont {{Cerd{\'a}-Dur{\'a}n}}(2010)}]{CerdaDuran2010}%
  \BibitemOpen
  \bibfield  {author} {\bibinfo {author} {\bibfnamefont {P.}~\bibnamefont {{Cerd{\'a}-Dur{\'a}n}}},\ }\bibfield  {title} {\bibinfo {title} {{Numerical viscosity in hydrodynamics simulations in general relativity}},\ }\href {https://doi.org/10.1088/0264-9381/27/20/205012} {\bibfield  {journal} {\bibinfo  {journal} {Classical and Quantum Gravity}\ }\textbf {\bibinfo {volume} {27}},\ \bibinfo {eid} {205012} (\bibinfo {year} {2010})},\ \Eprint {https://arxiv.org/abs/0912.1774} {arXiv:0912.1774 [astro-ph.SR]} \BibitemShut {NoStop}%
\bibitem [{\citenamefont {{Chabanov}}\ and\ \citenamefont {{Rezzolla}}(2025)}]{Chabanov2023b}%
  \BibitemOpen
  \bibfield  {author} {\bibinfo {author} {\bibfnamefont {M.}~\bibnamefont {{Chabanov}}}\ and\ \bibinfo {author} {\bibfnamefont {L.}~\bibnamefont {{Rezzolla}}},\ }\bibfield  {title} {\bibinfo {title} {{Numerical modeling of bulk viscosity in neutron stars}},\ }\href {https://doi.org/10.1103/PhysRevD.111.044074} {\bibfield  {journal} {\bibinfo  {journal} {\prd}\ }\textbf {\bibinfo {volume} {111}},\ \bibinfo {eid} {044074} (\bibinfo {year} {2025})},\ \Eprint {https://arxiv.org/abs/2311.13027} {arXiv:2311.13027 [gr-qc]} \BibitemShut {NoStop}%
\bibitem [{\citenamefont {Pereira}\ \emph {et~al.}(2018)\citenamefont {Pereira}, \citenamefont {Flores},\ and\ \citenamefont {Lugones}}]{pereira_phase_2018}%
  \BibitemOpen
  \bibfield  {author} {\bibinfo {author} {\bibfnamefont {J.~P.}\ \bibnamefont {Pereira}}, \bibinfo {author} {\bibfnamefont {C.~V.}\ \bibnamefont {Flores}},\ and\ \bibinfo {author} {\bibfnamefont {G.}~\bibnamefont {Lugones}},\ }\bibfield  {title} {\bibinfo {title} {Phase {{Transition Effects}} on the {{Dynamical Stability}} of {{Hybrid Neutron Stars}}},\ }\href {https://doi.org/10.3847/1538-4357/aabfbf} {\bibfield  {journal} {\bibinfo  {journal} {ApJ}\ }\textbf {\bibinfo {volume} {860}},\ \bibinfo {pages} {12} (\bibinfo {year} {2018})}\BibitemShut {NoStop}%
\bibitem [{\citenamefont {Gon{\c c}alves}\ \emph {et~al.}(2022)\citenamefont {Gon{\c c}alves}, \citenamefont {Jim{\'e}nez},\ and\ \citenamefont {Lazzari}}]{goncalves_fundamental-mode_2022}%
  \BibitemOpen
  \bibfield  {author} {\bibinfo {author} {\bibfnamefont {V.~P.}\ \bibnamefont {Gon{\c c}alves}}, \bibinfo {author} {\bibfnamefont {J.~C.}\ \bibnamefont {Jim{\'e}nez}},\ and\ \bibinfo {author} {\bibfnamefont {L.}~\bibnamefont {Lazzari}},\ }\bibfield  {title} {\bibinfo {title} {Fundamental-mode eigenfrequencies of neutral and charged twin neutron stars},\ }\href {https://doi.org/10.1140/epjc/s10052-022-11115-0} {\bibfield  {journal} {\bibinfo  {journal} {Eur. Phys. J. C}\ }\textbf {\bibinfo {volume} {82}},\ \bibinfo {pages} {1117} (\bibinfo {year} {2022})}\BibitemShut {NoStop}%
\bibitem [{\citenamefont {Rau}\ and\ \citenamefont {Sedrakian}(2023)}]{rau_two_2023}%
  \BibitemOpen
  \bibfield  {author} {\bibinfo {author} {\bibfnamefont {P.~B.}\ \bibnamefont {Rau}}\ and\ \bibinfo {author} {\bibfnamefont {A.}~\bibnamefont {Sedrakian}},\ }\bibfield  {title} {\bibinfo {title} {Two first-order phase transitions in hybrid compact stars: {{Higher-order}} multiplet stars, reaction modes, and intermediate conversion speeds},\ }\href {https://doi.org/10.1103/PhysRevD.107.103042} {\bibfield  {journal} {\bibinfo  {journal} {Phys. Rev. D}\ }\textbf {\bibinfo {volume} {107}},\ \bibinfo {pages} {103042} (\bibinfo {year} {2023})}\BibitemShut {NoStop}%
\bibitem [{\citenamefont {Harris}\ \emph {et~al.}(2020)\citenamefont {Harris} \emph {et~al.}}]{harris_array_2020}%
  \BibitemOpen
  \bibfield  {author} {\bibinfo {author} {\bibfnamefont {C.~R.}\ \bibnamefont {Harris}} \emph {et~al.},\ }\bibfield  {title} {\bibinfo {title} {Array programming with {{NumPy}}},\ }\href {https://doi.org/10.1038/s41586-020-2649-2} {\bibfield  {journal} {\bibinfo  {journal} {Nature}\ }\textbf {\bibinfo {volume} {585}},\ \bibinfo {pages} {357} (\bibinfo {year} {2020})}\BibitemShut {NoStop}%
\bibitem [{\citenamefont {Hunter}(2007)}]{hunter_matplotlib_2007}%
  \BibitemOpen
  \bibfield  {author} {\bibinfo {author} {\bibfnamefont {J.~D.}\ \bibnamefont {Hunter}},\ }\bibfield  {title} {\bibinfo {title} {Matplotlib: {{A 2D}} graphics environment},\ }\href {https://doi.org/10.1109/MCSE.2007.55} {\bibfield  {journal} {\bibinfo  {journal} {Computing in Science \& Engineering}\ }\textbf {\bibinfo {volume} {9}},\ \bibinfo {pages} {90} (\bibinfo {year} {2007})}\BibitemShut {NoStop}%
\bibitem [{\citenamefont {Kluyver}\ \emph {et~al.}(2016)\citenamefont {Kluyver} \emph {et~al.}}]{kluyver_jupyter_2016}%
  \BibitemOpen
  \bibfield  {author} {\bibinfo {author} {\bibfnamefont {T.}~\bibnamefont {Kluyver}} \emph {et~al.},\ }\bibfield  {title} {\bibinfo {title} {Jupyter {{Notebooks}} -- a publishing format for reproducible computational workflows},\ }in\ \href@noop {} {\emph {\bibinfo {booktitle} {Positioning and {{Power}} in {{Academic Publishing}}: {{Players}}, {{Agents}} and {{Agendas}}}}},\ \bibinfo {editor} {edited by\ \bibinfo {editor} {\bibfnamefont {F.}~\bibnamefont {Loizides}}\ and\ \bibinfo {editor} {\bibfnamefont {B.}~\bibnamefont {Schmidt}}}\ (\bibinfo  {publisher} {IOS Press},\ \bibinfo {year} {2016})\ pp.\ \bibinfo {pages} {87--90}\BibitemShut {NoStop}%
\bibitem [{Note3()}]{Note3}%
  \BibitemOpen
  \bibinfo {note} {We recall that the gravitational mass is numerically extracted at the outer boundary after a volume integral across the whole domain [see Eq.(4) in \cite {oconnor_open-source_2015}].}\BibitemShut {Stop}%
\bibitem [{\citenamefont {Kumar}\ \emph {et~al.}(2025)\citenamefont {Kumar}, \citenamefont {Karan}, \citenamefont {Verma}, \citenamefont {Mishra},\ and\ \citenamefont {Mallick}}]{kumar_modification_2025}%
  \BibitemOpen
  \bibfield  {author} {\bibinfo {author} {\bibfnamefont {D.}~\bibnamefont {Kumar}}, \bibinfo {author} {\bibfnamefont {A.}~\bibnamefont {Karan}}, \bibinfo {author} {\bibfnamefont {A.}~\bibnamefont {Verma}}, \bibinfo {author} {\bibfnamefont {H.}~\bibnamefont {Mishra}},\ and\ \bibinfo {author} {\bibfnamefont {R.}~\bibnamefont {Mallick}},\ }\bibfield  {title} {\bibinfo {title} {Modification of the universal relation between mass, radius, and nonradial f -mode oscillation in proto-neutron stars},\ }\href {https://doi.org/10.1103/PhysRevC.111.055805} {\bibfield  {journal} {\bibinfo  {journal} {Phys. Rev. C}\ }\textbf {\bibinfo {volume} {111}},\ \bibinfo {pages} {055805} (\bibinfo {year} {2025})}\BibitemShut {NoStop}%
\bibitem [{\citenamefont {Karan}()}]{karan_private}%
  \BibitemOpen
  \bibfield  {author} {\bibinfo {author} {\bibfnamefont {A.}~\bibnamefont {Karan}},\ }\href@noop {} {}\bibinfo {note} {(private communication)}\BibitemShut {NoStop}%
\bibitem [{\citenamefont {Read}\ \emph {et~al.}(2009)\citenamefont {Read}, \citenamefont {Lackey}, \citenamefont {Owen},\ and\ \citenamefont {Friedman}}]{read_constraints_2009}%
  \BibitemOpen
  \bibfield  {author} {\bibinfo {author} {\bibfnamefont {J.~S.}\ \bibnamefont {Read}}, \bibinfo {author} {\bibfnamefont {B.~D.}\ \bibnamefont {Lackey}}, \bibinfo {author} {\bibfnamefont {B.~J.}\ \bibnamefont {Owen}},\ and\ \bibinfo {author} {\bibfnamefont {J.~L.}\ \bibnamefont {Friedman}},\ }\bibfield  {title} {\bibinfo {title} {Constraints on a phenomenologically parametrized neutron-star equation of state},\ }\href {https://doi.org/10.1103/PhysRevD.79.124032} {\bibfield  {journal} {\bibinfo  {journal} {Phys. Rev. D}\ }\textbf {\bibinfo {volume} {79}},\ \bibinfo {pages} {124032} (\bibinfo {year} {2009})}\BibitemShut {NoStop}%
\bibitem [{\citenamefont {{Timmes}}\ \emph {et~al.}(1996)\citenamefont {{Timmes}}, \citenamefont {{Woosley}},\ and\ \citenamefont {{Weaver}}}]{Timmes1996}%
  \BibitemOpen
  \bibfield  {author} {\bibinfo {author} {\bibfnamefont {F.~X.}\ \bibnamefont {{Timmes}}}, \bibinfo {author} {\bibfnamefont {S.~E.}\ \bibnamefont {{Woosley}}},\ and\ \bibinfo {author} {\bibfnamefont {T.~A.}\ \bibnamefont {{Weaver}}},\ }\bibfield  {title} {\bibinfo {title} {{The Neutron Star and Black Hole Initial Mass Function}},\ }\href {https://doi.org/10.1086/176778} {\bibfield  {journal} {\bibinfo  {journal} {Astrophys. J.}\ }\textbf {\bibinfo {volume} {457}},\ \bibinfo {pages} {834} (\bibinfo {year} {1996})},\ \Eprint {https://arxiv.org/abs/astro-ph/9510136} {astro-ph/9510136} \BibitemShut {NoStop}%
\bibitem [{\citenamefont {Baiotti}\ \emph {et~al.}(2005)\citenamefont {Baiotti}, \citenamefont {Hawke}, \citenamefont {Montero}, \citenamefont {L{\"o}ffler}, \citenamefont {Rezzolla}, \citenamefont {Stergioulas}, \citenamefont {Font},\ and\ \citenamefont {Seidel}}]{baiotti_three-dimensional_2005}%
  \BibitemOpen
  \bibfield  {author} {\bibinfo {author} {\bibfnamefont {L.}~\bibnamefont {Baiotti}}, \bibinfo {author} {\bibfnamefont {I.}~\bibnamefont {Hawke}}, \bibinfo {author} {\bibfnamefont {P.~J.}\ \bibnamefont {Montero}}, \bibinfo {author} {\bibfnamefont {F.}~\bibnamefont {L{\"o}ffler}}, \bibinfo {author} {\bibfnamefont {L.}~\bibnamefont {Rezzolla}}, \bibinfo {author} {\bibfnamefont {N.}~\bibnamefont {Stergioulas}}, \bibinfo {author} {\bibfnamefont {J.~A.}\ \bibnamefont {Font}},\ and\ \bibinfo {author} {\bibfnamefont {E.}~\bibnamefont {Seidel}},\ }\bibfield  {title} {\bibinfo {title} {Three-dimensional relativistic simulations of rotating neutron-star collapse to a {{Kerr}} black hole},\ }\href {https://doi.org/10.1103/PhysRevD.71.024035} {\bibfield  {journal} {\bibinfo  {journal} {Phys. Rev. D}\ }\textbf {\bibinfo {volume} {71}},\ \bibinfo {pages} {024035} (\bibinfo {year} {2005})}\BibitemShut {NoStop}%
\bibitem [{\citenamefont {O'Connor}(2015)}]{oconnor_open-source_2015}%
  \BibitemOpen
  \bibfield  {author} {\bibinfo {author} {\bibfnamefont {E.}~\bibnamefont {O'Connor}},\ }\bibfield  {title} {\bibinfo {title} {{{An open-source neutrino radiation hydrodynamics code for core-collapse supernovae}}},\ }\href {https://doi.org/10.1088/0067-0049/219/2/24} {\bibfield  {journal} {\bibinfo  {journal} {ApJS}\ }\textbf {\bibinfo {volume} {219}},\ \bibinfo {pages} {24} (\bibinfo {year} {2015})}\BibitemShut {NoStop}%
\end{thebibliography}%

\end{document}